\documentclass[a4paper,fleqn,usenatbib]{mnras}

\usepackage[T1]{fontenc}
\usepackage{ae,aecompl}

\usepackage{ulem}
\usepackage{graphicx}

\usepackage{txfonts}

\usepackage{hyperref}
\hypersetup{colorlinks=true, linkcolor=blue, citecolor=blue, urlcolor=blue}

\usepackage{color}
\usepackage{amstext}
\usepackage{multirow}
\usepackage{todonotes}

\newcommand*\lephare{L\textsc{e} P\textsc{hare}}
\newcommand{\sextractor}{{\sc SExtractor}}

\newcommand{\newu}{$u$}
\newcommand{\oldu}{$u^*$}
\newcommand{\sqdeg}{deg$^{2}$}

\newcommand{\Sing}{{\mathcal{S}ing}}
\newcommand{\Doub}{{\mathcal{D}oub}}

\usepackage{soul}

\title[UV and U-band LFs from CLAUDS and HSC-SSP]{UV \& U-band luminosity functions from CLAUDS and HSC-SSP -- I. Using four million galaxies to simultaneously constrain the very faint and bright regimes to z $\sim$ 3 
}

\author[T. Moutard et al.]
{Thibaud Moutard,$^{1}$\thanks{thibaud.moutard@lilo.org}
Marcin Sawicki,$^{1,2}$\thanks{Canada Research Chair}
St\'ephane Arnouts,$^{3}$
Anneya Golob,$^{1}$ 
\newauthor 
Jean Coupon,$^{4}$
Olivier Ilbert,$^{3}$
Xiaohu Yang,$^{5}$
Stephen Gwyn$^{2}$
\newauthor 
\\
$^{1}$Department of Astronomy \& Physics and Institute for Computational Astrophysics, Saint Mary's University, 923 Robie Street, Halifax,  \\
~~Nova Scotia, B3H 3C3, Canada \\
$^{2}$NRC Herzberg Astronomy and Astrophysics, 5071 West Saanich Road, Victoria, BC V9E 2E7, Canada \\
$^{3}$Aix Marseille Universit\'e, CNRS, LAM - Laboratoire d'Astrophysique de Marseille, 38 rue F. Joliot-Curie, F-13388, Marseille, France \\
$^{4}$Astronomical Observatory of the University of Geneva, ch. d'Ecogia 16, 1290 Versoix, Switzerland \\
$^{5}$Department of Astronomy, Shanghai Jiao Tong University, Dongchuan RD 800, 200240 Shanghai, China
}

\date{Accepted XXX. Received YYY; in original form ZZZ}

\pubyear{2019}

\begin{document}
\label{firstpage}
\pagerange{\pageref{firstpage}--\pageref{lastpage}}
\maketitle



\begin{abstract}
We constrain the rest-frame FUV (1546\AA), NUV (2345\AA) and U-band (3690\AA) luminosity functions (LFs) and luminosity densities (LDs) with unprecedented precision from $z\sim0.2$ to $z\sim3$ (FUV, NUV) and $z\sim2$ (U-band). Our sample of over 4.3 million galaxies, selected from the CFHT Large Area $U$-band Deep Survey (CLAUDS) and  HyperSuprime-Cam Subaru Strategic Program (HSC-SSP) data lets us probe the very faint regime (down to $M_\mathrm{FUV},M_\mathrm{NUV},M_\mathrm{U} \simeq -15$ at low redshift)  while simultaneously detecting very rare galaxies at the bright end down to comoving densities $<10^{-5}$ Mpc$^{-3}$. Our FUV and NUV LFs are well fitted by single Schechter functions, with faint-end slopes that are very stable up to $z\sim2$. We confirm, but self-consistently and with much better precision than previous studies, that the LDs at all three wavelengths increase rapidly with lookback time to $z\sim1$, and then much more slowly at $1<z<2$--$3$. Evolution of the FUV and NUV LFs and LDs at $z<1$ is driven almost entirely by the fading of the characteristic magnitude, $M^\star_{UV}$, while at $z>1$ it is due to the evolution of both $M^\star_{UV}$ and the characteristic number density $\phi^\star_{UV}$.  In contrast, the U-band LF has an excess of faint galaxies and is fitted with a double-Schechter form; $M^\star_\mathrm{U}$, both $\phi^\star_\mathrm{U}$ components, and the bright-end slope evolve throughout $0.2<z<2$, while the faint-end slope is constant over at least the measurable $0.05<z<0.6$. We present tables of our Schechter parameters and LD measurements that can be used for testing theoretical galaxy evolution models and forecasting future observations.
\end{abstract}






\section{Introduction}
\label{introduction}

The galaxy luminosity function (LF) and its redshift evolution is one of the most fundamental ways to characterize the galaxy population. It provides a direct probe of the hierarchical framework of galaxy formation. Defined by $\phi(L)\ dL$ as the comoving number density of galaxies with luminosity between $L$ and $L + dL$, the LF is a wavelength-dependent measurement that gives a direct test on the modelling of the baryonic physics such as star formation activity, dust attenuation and feedback processes.
The present paper is concerned with galaxy LFs at rest-frame ultra-violet (UV: $\lambda = 1000-3000$~\AA) and u ($\lambda = 3000-4000$~\AA) wavelengths.
In this wavelength regime, light in star-forming galaxies is thought to be primarily produced by short-lived massive stars.
For this reason, the evolution of the UV LF has historically been used as a probe of the evolution of star-forming activity in the galaxy population.  

Similarly, the UV luminosity density ($\rho_{UV}$) -- which is the luminosity-weighted integral of the LF, $\int L \times \phi(L)\ dL$ -- is a direct measurement of the unobscured cosmic star formation density (SFRD, $\rho_{SFR})$ and its evolution with redshift, giving us a sketch of the cosmic star formation history.  
At $0\lesssim z\lesssim1$, this was first done by  \citet{Lilly1996}, with UV LFs measurements from spectroscopic samples with optically selected sources \citep[][]{Lilly1995}.
At higher redshift, a lower limit on $\rho_{SFR}$ based on Lyman-break galaxies (LBGs) was determined by \citet{Madau1996} who summed up the UV light from $U$- and $B$-band dropouts detected in the Hubble Deep Field \citep[HDF,][]{Williams1996}, while \citet{Sawicki1997} presented the first measurement of $\rho_{SFR}$ between $z = 1$ and $z\sim3.5$ by making use of photometric redshifts.
The realization that significant fractions of UV photons are prevented from escaping from high-$z$ star-forming galaxies by interstellar dust \citep[e.g.,][]{Meurer1997, Sawicki1998} forced dust corrections to be subsequently applied to this  $\rho_{UV} \rightarrow \rho_{SFR}$ conversion method. 

Subsequently, the GALEX satellite  \citep{Martin2005b} allowed first measurements of the unobscured UV LF at $z\sim 0$ \citep{Wyder2005,Budavari2005} and out to $z\sim1.2$ \citep{Arnouts2005}. The later provided the first UV LF measurements over the entire redshift range $0\le z\le 3.5$ by combining spectroscopically selected GALEX sources and photometric redshifts of optically selected sources at high redshift in the HDFs. \citet{Schiminovich2005} used those UV-LFs to estimate the evolution of the UV luminosity density ($\rho_{UV}$) and of the cosmic star formation rate density ($\rho_{SFR}$) after accounting for typical UV attenuation due to interstellar dust in star-forming galaxies.

SFRD measurements made at infra-red or sub-millimetre wavelengths -- which measure the stellar energy re-radiated by interstellar dust and thus obviate the needs for dust corrections -- can provide a complementary picture to that gleaned from the UV.  Although such measurements have been possible for some time for high-$z$ galaxies \citep[e.g.,][]{
Hughes1998,
Chapman2005, 
Magnelli2013, 
Gruppioni2013, Goto2019}, 
they do not yet provide significant insights at very high redshifts ($z \ga 6$), nor for low-mass galaxies which have low SFRs and low dust content \citep[e.g.,][]{Bouwens2009, Bouwens2012, Sawicki2012}. 

Consequently,  UV LF measurements allow us the only self-consistent way to study the evolution of the galaxy population and of the SFRD {\it at a constant rest-frame wavelenght} across the entire redshift range over which galaxies are currently known to exist, $z=0\sim10$.   Similarly, UV measurements let us reach galaxies that are too faint to be observed by infra-red and sub-mm surveys.
This explains why UV LFs have continuously been used for estimating the evolution of the cosmic star formation rate density over the last two decades \citep[e.g.,][and many others]{
Steidel999, 
Ouchi2004, 
SawickiThompson2006b,
Dahlen2007,
Iwata2007,
Reddy2009, 
vanderBurg2010,
Cucciati2012,
Sawicki2012, 
McLure2013,
MadauDickinson2014,
Bouwens2015a, 
Bouwens2016,
Ono2018,
Khusanova2019}.

The advent of multi-wavelength datasets that contain flux measurements at a great many wavelengths (sometime as many as several dozen --- e.g.,  \citealt{Laigle2016}) allow the estimation of physical quantities for each galaxy, such as its stellar mass, and the construction of related global descriptors, such as the galaxy stellar mass functions (SMFs) and stellar mass densities (SMDs, $\rho_{M_\star}$) -- e.g., \cite{Ilbert2013, Muzzin2013, Moutard2016b, Davidzon2017}.  
Such "physical" measurements are an extremely powerful tool to help us understand galaxy evolution, but they suffer from some important limitations: they rely heavily on the assumptions that underpin stellar population synthesis models \citep[e.g.,][]{BC2003, Maraston2005}, and the spectral energy distribution (SED) -fitting technique that's used for physical parameter estimation (e.g., \citealt{Sawicki1998, Papovich2001}; see \citealt{Conroy2013} for a review).   Consequently, the fidelity of the physical parameter estimates continues to be challenged by studies that show that biases may exist in commonly-used approaches: for example, different galaxy star formation histories \citep[e.g.,][]{Leja2019}, the assumed  stellar initial mass function \citep[IMF; e.g.,][]{Salpeter1955,Chabrier2003}, the common assumption that dust acts as a uniform foreground screen \citep[see, e.g.,][]{Mitchell2013}, or the treatment of individual galaxies as consisting of spatially-homogeneous stellar populations \citep[e.g.,][]{Sorba2015, Sorba2018}, can influence the inferred stellar masses and -- consequently -- SMFs and SMDs. While such "physical" measurements are a powerful tool to help us understand galaxy evolution, model-independent measurements, such as LFs, are therefore an essential complement.

One example of direct applications of the LFs is to calibrate or validate galaxy formation models \citep[e.g.,][]{KitzbichlerWhite2007, Lacey2011, Somerville2012, Henriques2013, Lacey2016, Sharma2016}, since in using the directly-measured quantity (i.e., the LF), the modeller has full control over the comparison process, rather than relying on assumptions made by the observational papers. A related use of UV LFs is in the forecasting of future observations \citep[e.g.,][]{Williams2018, Maseda2019}.  Finally, because UV LFs probe the galaxy population at wavelengths close to those which ionize hydrogen, UV LFs are used in work that aims to assess the contribution of different types of objects to reionizing the Universe, or to maintaining it in its ionized state \citep[e.g.,][]{Inoue2006, SawickiThompson2006b, Bouwens2015b, Ishigaki2018, Iwata2019}. 

For these reasons it is important that we have the best possible measurements of the UV LFs over a wide redshift range of cosmic history. Although the situation has improved dramatically from the early days of the Hubble Deep Field, even the largest studies to date are still based on relatively small fields, such as the COSMOS field \citep{Scoville2007}, and are thus susceptible to cosmic variance and poor statistics, particularly at the bright end.  With new data that we now have in hand, we can do better.  In this paper we therefore set out to provide a state-of-the-art measurement of the rest-frame FUV (1546~\AA), NUV (2345~\AA), and U-band (3690~\AA) luminosity functions using two overlapping and complementary cutting-edge surveys:  the recently-completed Canada-France-Hawaii Telescope (CFHT) Large Area U-band Deep Survey \citep[CLAUDS,][]{Sawicki2019} and the ongoing HyperSuprime-Cam Subaru Strategic Program \citep[HSC-SSP,][]{Aihara2018overview}.  Together, these two surveys probe the Universe to an unprecedented combination of area and depth, as described in Sec.~\ref{sect_data} and allow us to produce the most statistically-significant measurements of the UV LFs that are also essentially free of cosmic variance. 

This paper focuses on providing reference measurements of the rest-frame FUV, NUV, and U-band LFs based on these state-of-the-art surveys, notably to serve as a basis for making observational forecasts and validating theoretical models. 
We postpone more physically-motivated interpretations to future work (see companion paper, T.~Moutard et al.~in prep.). 

Throughout this paper, we use the standard cosmology ($\Omega_m~=~0.3$, $\Omega_\Lambda~=~0.7$ with $H_{\rm0}~=~70$~km~s$^{-1}$~Mpc$^{-1}$).
Magnitudes are given in the AB system \citep{Oke1974}.

\section{Galaxy sample}
\label{sect_gal_sample}

\subsection{Data}
\label{sect_data}

\begin{figure}
\center
\includegraphics[width=\hsize, trim = 1.1cm 0.5cm 2.5cm 2.1cm, clip]{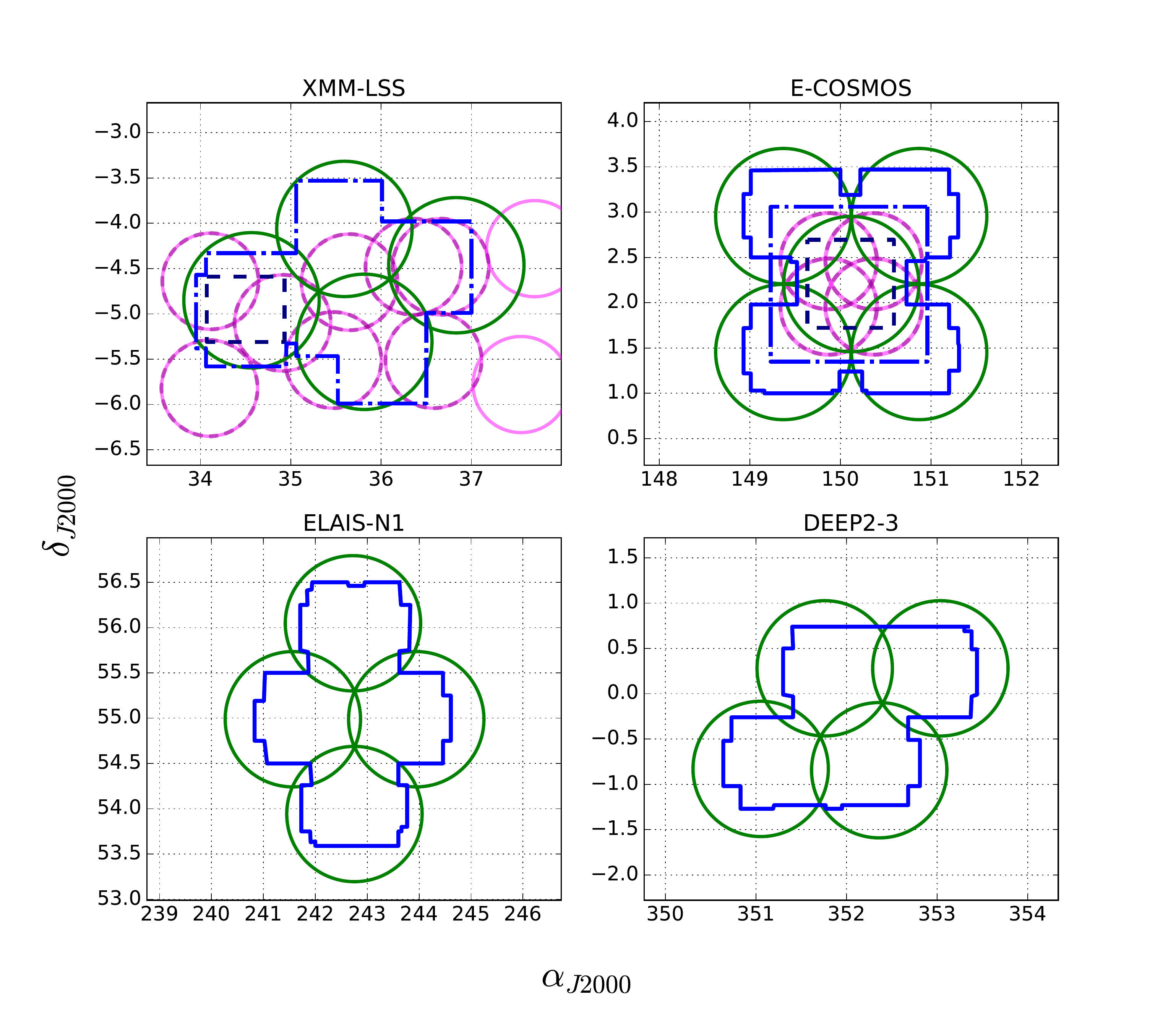}
\caption{Footprints of the CLAUDS Deep layer $u$- and $u^*$-bands (blue solid and long-dash-dotted outlines, respectively) and HSC-SSP $grizy$ data (green circles).  The area where the two surveys overlap totals 18.60~\sqdeg\ and reaches a median depth of $U = 27.1$, and a minimum depth of $U \geq 26.8$ (5$\sigma$ in 2\arcsec apertures). The CLAUDS Ultra-Deep layer (dark-blue dashed outlines) covers 1.55~\sqdeg\ over the XMM-LSS and E-COSMOS fields down to a minimum depth of $U \geq 27.5$. GALEX $FUV$ (dark magenta dashed circles) and $NUV$ (light magenta solid circles) observations overlap over 6 and 7 ~\sqdeg, respectively, down to a median depth of $FUV,NUV \sim 25$.
\label{fig_footprint}  }
\end{figure}

This study uses the $U+grizy$ data from the Canada-France-Hawaii Telescope (CFHT) Large Area U-band Deep Survey (CLAUDS) and the HyperSuprime-Cam Subary Strategic Program (HSC-SSP). These surveys are described in detail in \citet[][CLAUDS]{Sawicki2019} and in \citet[][and references therein; HSC-SSP]{Aihara2018overview}, and the procedures for merging the datasets are described in \cite{Sawicki2019}.  Consequently, here we give only a summary of the key details.

The CLAUDS and HSC-SSP imaging data overlap over four well-studied fields, namely E-COSMOS, ELAIS-N1, DEEP2-3, and XMM-LSS, each spanning $\sim$4--6~\sqdeg. The $U$-band data cover 18.60~\sqdeg\ to a depth of $U_{AB}$=27.1 (5$\sigma$ in 2\arcsec apertures), with selected ultra-deep sub-areas within the E-COSMOS and XMM-LSS fields that cover 1.36~\sqdeg\ to a depth of $U$=27.7 (5$\sigma$ in 2\arcsec apertures).  CLAUDS $U$-band data were obtained in two somewhat different CFHT/MegaCam filters: data in the ELAIS-N2 and DEEP2-3 fields were taken with the new \newu\ filter, while those in XMM-LSS were taken with the older \oldu\ filter.  The E-COSMOS field contains data in the \newu\ filter except in the central region where both \newu\ and \oldu\ data overlap. The \newu\ and \oldu\ data are kept separate, even in areas where they overlap. The image quality of the CLAUDS data is excellent, with median seeing of 0.92\arcsec.  For the details of CLAUDS data see \cite{Sawicki2019}.

The HSC-SSP project \citep{Aihara2018overview} provides deep Subary/HSC imaging in the $grizy$ wavebands in the same fields imaged by CLAUDS. Here we use images from the S16A internal HSC-SSP data release that are deeper than the HSC-SSP public data release 1 \citep[PDR1,][]{Aihara2018dr1} with depths of $g_{AB} \sim 26.6$, $r_{AB} \sim 26.1$, $i_{AB} \sim 25.7$, $z_{AB} \sim 25.1$ and $y_{AB} \sim 24.2$ (5$\sigma$ in 2\arcsec apertures), though not as deep as those from the very recent PDR2 \citep{Aihara2019} .  Seeing in the HSC-SSP varies from band to band, with the $i$-band providing the sharpest images ($\sim$0.62\arcsec); in all bands, the seeing in the HSC images is even better than the (excellent) seeing in the CLAUDS $U$-band data. 

Figure~\ref{fig_footprint} shows the overlap of the CLAUDS (black) and HSC-SSP (green) footprints. The footprints of the the deep HSC observations are somewhat larger than those of the CLAUDS data, so the area of overlap is dictated by the extent of the CLAUDS data, i.e., 18.29~\sqdeg\ after the masking of areas around bright stars. Our survey contains two layers of different depths: 
\begin{itemize}
\item the Deep layer covers a total area of 18.29 deg$^2$  with $U\geq26.8$, $g\geq26.5$, $r\geq26.1$, $i\geq25.7$, $z\geq25.1$ and $y\geq24.2$, respectively; 
\item while the Ultra-Deep layer covers an area of 1.54 deg$^2$ with $U\geq27.5$, $g\geq27.1$, $r\geq26.9$, $i\geq26.6$, $z\geq26.3$ and $y\geq25.0$, respectively.
\end{itemize}

We use the \sextractor-based multi-band catalog described in \cite{Sawicki2019}. For object detection, this uses the signal-to-noise image, $\Sigma\rm SNR$ constructed from all the available $uu^*grizy$ images as 
\begin{equation}
\Sigma{\mathrm{SNR}}=\sum_{i=1}^{N}\left(\frac{f_{i}-\mu_{i}}{\sigma_{i}}\right),\label{eq:combUgrizy}
\end{equation}
where $f_i$ is the flux in each pixel, $\sigma_i$ is the RMS width of the background sky distribution, and $\mu_i$ is its mean. Here the index $i$ runs over MegaCam bands \newu\ or \oldu\ (or both, where available --- i.e.,  in the central area of E-COSMOS) as well as the HSC bands $grizy$.  Once the \sextractor\ software \citep{Bertin1996} has detected objects in the $\Sigma{\mathrm{SNR}}$ image, the multiband catalog is then created by running \sextractor\ in dual image mode, with various measurements recorded for each object, including positions, fluxes (in Kron, isophotal, and fixed-radius circular apertures), fiducial radii, ellipticities, position angles, and central surface brightnesses. 
For more details see \cite{Sawicki2019}  and A.~Golob et al.\ (submitted to MNRAS).  Note that the CLAUDS $U$-band images are as deep or deeper than the HSC-SSP S16A images we used and consequently our catalog is not expected to be biased against $U$-faint objects. 

Small apertures are known to provide less noisy colours and therefore an improved photometric redshift accuracy than total Kron-like \citep{Kron1980} apertures \citep{Sawicki1997, Hildebrandt2012, Moutard2016a, Moutard2016b}.  At the same time, total fluxes are needed for deriving galaxy physical properties. Following the approach of \citet{Moutard2016a}, the final magnitudes $m_\textsc{final}$ of each source are produced by rescaling isophotal magnitudes $m_\textsc{iso}$ to the Kron-like magnitudes $m_\textsc{auto}$. To preserve the colours based on isophotal apertures, a mean rescaling factor $\delta m$ is applied in each filter $f$: 
\begin{equation}
m_{\textsc{final},f} =  m_{\textsc{iso},f} + \delta m 
\end{equation}
with $\delta m$ defined as 
\begin{equation}
\delta m =  \frac{  \sum_{f}  (m_{\textsc{auto},f}- m_{\textsc{iso},f})  \times w_f}{ \sum_{f}  w_f} 
\end{equation}
for $f = u,u^*,g,r,i,z,y$ and where the weights $w_f$ are simply defined from $\sigma_\textsc{iso}$ and $\sigma_\textsc{auto}$, the photometric uncertainties on $m_\textsc{iso}$ and $m_\textsc{auto}$, with $w_f = 1 /(\sigma_{\textsc{auto},f}^2+\sigma_{\textsc{iso},f}^2)$.

To properly constrain the FUV and NUV luminosities at low redshift, we complemented our photometric dataset with FUV (135-175 nm) and NUV (170-275 nm) observations from the GALEX satellite \citep{Martin2005b}. Both in the XMM-LSS and E-COSMOS fields, the GALEX observations we used were reduced with the {\sc EMphot} code \citep{Guillaume2006, Conseil2011} dedicated to extract UV photometry by using the CFHTLS (T0007) $u^*$-band detections as a priors down to \oldu$\sim$25. Consequently, the astrometry of the resulting GALEX photometry is that of the CFHTLS,  which enabled a straightforward position matching with our photometric dataset (with 0.5\arcsec\ tolerance).

\subsection{Galaxy identification and photometric redshift estimation}
\label{sect_photoz}

\begin{figure}
\center
\includegraphics[width=\hsize]{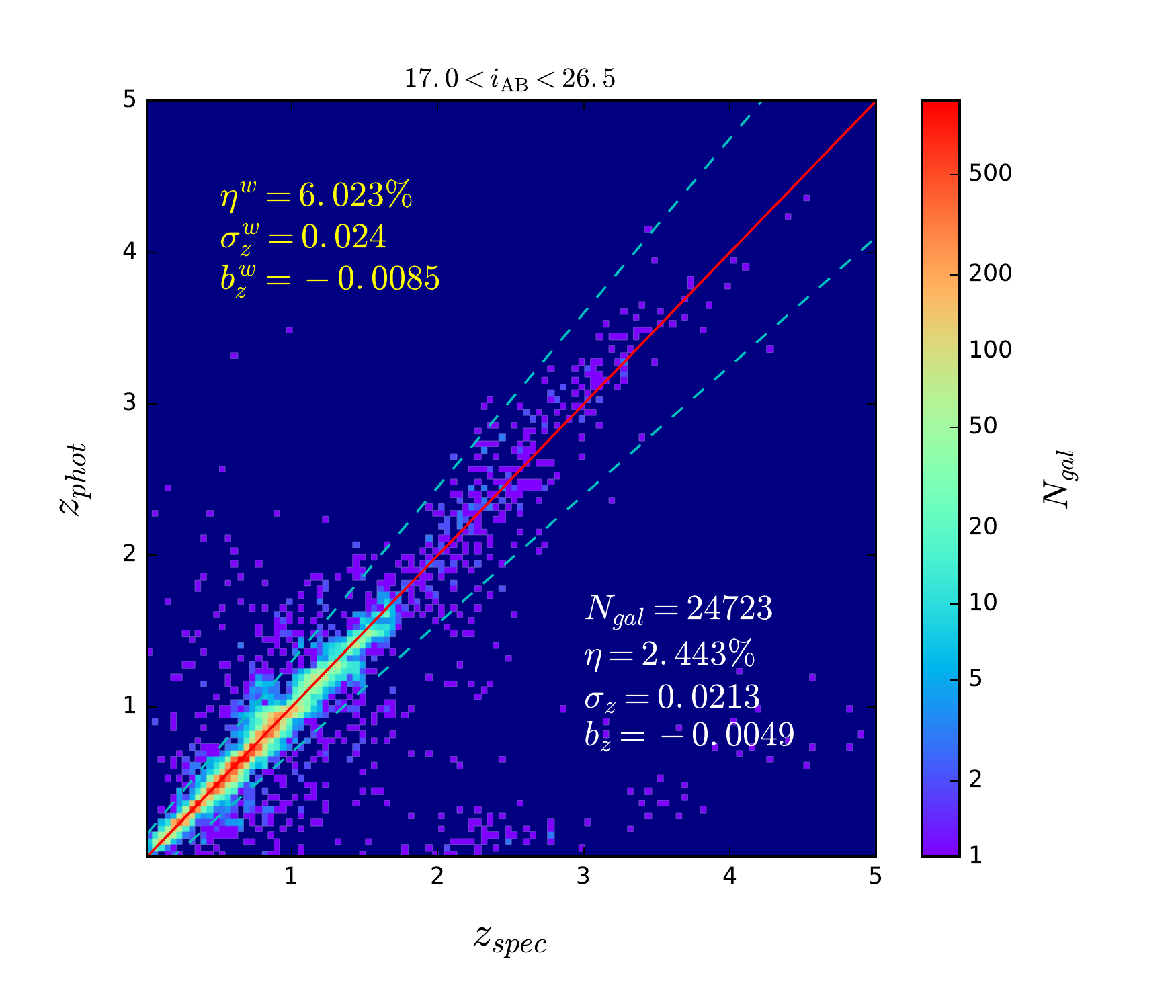}
\caption{Comparison of our photometric redshifts with spectroscopic redshifts from \citet[][]{bra13, com15, lef13, kri15, lil07, mas17, mas19, mcl13, sco18, sil15, tas17}.  The red diagonal line shows equality (perfect match) and the dashed blue lines define outliers. 
The total number of galaxy spectroscopic redshifts and the usual photo-z accuracy estimators (outlier rate $\eta$, scatter $\sigma_z$ and bias $b_z$) are reported in the lower-right corner, while corresponding i-band weighted estimators are reported in the upper-left corner or the figure.
\label{fig_photoz}  }
\end{figure}

To identify and remove  foreground Galactic stars we use the machine-learning method and results of A.~Golob et al.\ (submitted to MNRAS).  This method uses both photometric and morphological information to classify objects as stars or galaxies.  In more detail, we use HST morphological object classification in the COSMOS field from \cite{Leauthaud2007} to train a gradient boosted tree (GBT) machine classifier to classify objects based on their CLAUDS+HSC-SSP  $Ugrizy$ magnitudes, colours, central surface brightnesses, and effective radii.  Because the method uses photometric information, it does well even for faint objects where morphologies from ground-based imaging are ambiguous. Having trained the GBT machine classifier, we use it to remove from our sample all objects for which the classifier returned a value greater than 0.89. Doing so, we discarded $\sim$7.4\% of the sources as stars. See A.~Golob et al.\ (submitted to MNRAS)  for details of the method and its application to our CLAUDS+HSC-SSP dataset. 

Our photometric redshifts are computed using a hybrid approach that combines a nearest-neighbours  machine-learning method (hereafter kNN; A.~Golob, in preparation; see also \citealt{Sawicki2019}) with the template-fitting code \lephare~ \citep{Arnouts2002, Ilbert2006}. 

The kNN method uses the 30-band COSMOS photometric redshifts from \cite{Laigle2016} as a training set.  For each object in our catalog it identifies 50 nearest neighbours in colour space and then fits a weighted Gaussian kernel density estimator (KDE), with each neighbour's redshift weighted by $(d_\mathrm{NN}\times \Delta z)^{-1}$; here $d_\mathrm{NN}$ is the Euclidean distance in colour space to the object under consideration, and $\Delta z$ is the width of the 68\% confidence interval of the neighbour's redshift in the \cite{Laigle2016} catalog.  We find that this method gives very good results on average (low scatter, $\sigma_z$, and bias, $b_z$) but suffers from more outliers than we would wish. 

Following \citet{Moutard2016a}, \lephare~ photometric redshifts were computed by making use of the template library of \citet{Coupon2015}, while considering four extinction laws with a reddening excess  E(B-V) $\leq$ 0.3, as described in \citet{Ilbert2009}. In addition,  as described in  \citet{Ilbert2006}, \lephare~  tracked down and corrected for any systematic difference between the photometry and the predicted magnitudes in each band, while using the known $N(z)$ at given apparent magnitude as a prior to avoid catastrophic failures. \lephare~ is thereby naturally well suited to take care of any fluctuation of the absolute calibration from field to field and to deal with the confusion between spectrum breaks in the absence of near-infrared observations.

Our hybrid photometric redshift method combines the outputs from the kNN method and \lephare\ as follows. 
We flag outliers in the kNN photo-$z$ catalogue and then replace their photometric redshift values with those from the \lephare\ template-fitting code.  Outliers are identified and flagged by comparison of the kNN redshift, $z_{KDE}$, with  the \lephare\ redshift, $z_{LPh}$. Specifically, when the threshold of $\Delta z_{phot} =  | z_{KDE} - z_{LPh} | / \sqrt{2} = 0.15 \times (1+\overline{z})$ is exceeded, with $\overline{z} = (z_{KDE} + z_{LPh}) /2$, we adopt $z_{LPh}$; otherwise we use $z_{KDE}$. Doing so, we notably reduced by half the number of photo-z outliers that are due to the confusion between the Lyman and Balmer breaks.

Figure~\ref{fig_photoz} shows the comparison of our hybrid photometric redshifts with a large sample of spectroscopic redshifts compiled from the literature. Overall, the hybrid photo-z quality is found to be very good within the ranges of redshift and magnitude we explore in our analysis, namely, up to $z = 3.5$ and for observed magnitudes $17.0<i_\mathrm{AB}<26.5$, with a scatter\footnote{Using the NMAD (normalized median absolute deviation) to define the scatter, $\sigma_z = 1.48 \times \mathrm{median}\left(~\frac{|z_{phot}-z_{spec}|}{1+z_{spec}}~\right)$.} of $\sigma_z = 0.0213$, a median bias\footnote{We simply define the photo-z bias as $b_z = \frac{z_{phot}-z_{spec}}{1+z_{spec}}$.} of $b_z = -0.0049$, and an outlier rate\footnote{$\eta$ is defined as the percentage of galaxies with $\frac{|z_{phot}-z_{spec}|}{1+z_{spec}}>0.15$.} of $\eta = 2.443\%$.
While the spectroscopic sample we assembled combines many surveys, which makes it as representative as possible, it is much brighter (and bluer) than our photometric sample. In order to account for this effect, we followed the approach of \citet{Moutard2016b} and weighted the photo-z accuracy estimators with respect to the i-band distribution of the photometric sample. Using this approach, we found weighted scatter of $\sigma^w_z = 0.024$, weighted median bias of $b^w_z = -0.0085$ and weighted outlier rate of $\eta^w = 6.023\%$ at $17.0 < i_\mathrm{AB} < 26.5$. These measurements confirmed the reliability of our hybrid photometric redshifts, which we use for the rest of the analysis that follows.

\subsection{Galaxy physical parameters}
\label{sect_phys_param}

\subsubsection{Physical parameters and absolute magnitudes}

Our procedure for estimating galaxy rest-frame FUV, NUV, and U-band magnitudes interpolates (or, in some cases -- extrapolates) from the observed photometry using spectral models fitted to the photometry.  We therefore describe these models (which also yield some physical parameters for our galaxies, such as their stellar masses) before moving on to describe the estimation of rest-frame magnitudes. 

Absolute magnitudes and other physical parameters (stellar mass, star formation rate, etc.) were derived with the template-fitting code \lephare, after fixing the redshift to its best estimate (i.e., our hybrid photometric redshifts -- see Sec. \ref{sect_photoz}). Following \citet{Moutard2016b}, we made use of the stellar population synthesis models of \citet{BC2003} and considered two metallicities, exponentially declining star formation histories that follow  $\tau^{-1} e^{-t/\tau}$ (as described in \citealt{Ilbert2013}), and three extinction laws with a maximum dust reddening of E(B-V) $=$ 0.5. Finally, we imposed a low extinction for low-SFR galaxies and the emission-line contribution was taken into account \citep[for more details see][]{Moutard2016b}.

We computed FUV, NUV and U-band absolute magnitudes by adopting the procedure followed by \citet[][]{Ilbert2005} to minimise the dependence of the absolute magnitudes to the template library.  Specifically, to minimize the k-correction term, the absolute magnitude in a given passband centered on $\lambda^0$  was derived from the observed magnitude in the filter passband that was the closest from $\lambda^0 \ \times \ (1+z)$, except -- to avoid measurements that are too noisy -- when the apparent magnitude had an error above 0.3 mag. Moreover, all rest-frame magnitudes were derived with two different template libraries \citep[][]{BC2003, Coupon2015}, which allowed us to verify that no significant systematic uncertainties were introduced by the choice of template library.

\subsubsection{Absolute magnitude error budget} 
\label{sect_absmag_err}

Of particular importance for our analysis are the uncertainties affecting the absolute magnitudes for which we measure the luminosity function.  The first source of uncertainty is the fitting error, $\sigma_{fit}$, which comes from the propagation of the photon noise. The fitting error contribution is directly estimated from the 1$\sigma$ dispersion of absolute magnitudes derived from observed photometry perturbed with associated errors.  The second source of uncertainty on the magnitude, $\sigma_{M,z}$, comes from the photometric redshift uncertainty.  One way to estimate its effect is to compare the absolute magnitudes derived with photometric and spectroscopic redshifts. While limited by the completeness of the spectroscopic sample, it is the most comprehensive estimate of the photo-z error contribution we have access to. The last source of uncertainty we considered, $\sigma_\mathit{SED}$,  comes from the choice of template library used to derive absolute magnitudes. The  $\sigma_\mathit{SED}$ uncertainties are expected to be negligible when the k-correction is small, which we ensured by limiting our analysis to a redshift range where the rest-frame emission is observed in one of our filters. To estimate $\sigma_\mathit{SED}$, we compared the absolute magnitudes derived from the empirical SED library \citep{Coupon2015}, $M^\textsc{emp}$, and from the stellar population synthesis models library \citep{BC2003}, $M^\textsc{sps}$, and take $\sigma_\mathit{SED} = | M^\textsc{emp}-M^\textsc{sps} | / \sqrt{2}$. The total absolute magnitude error is then given by
\begin{equation}
\sigma_{M} = \sqrt{ ~\sigma_{fit}^2 + \sigma_{M,z}^2~ +  ~\sigma_\mathit{SED}^2} ~.
\label{eq_err_MABS}
\end{equation}


\section{Results}
\label{sect_results}

\subsection{FUV, NUV and U-band luminosity functions}

\subsubsection{Completeness limits and wedding cake approach}
\label{sect_complim}

Following an approach similar to that in \citet{Pozzetti2010}, we based our estimate of the luminosity (or absolute magnitude) completeness limit on the  distribution of the faintest luminosity (or absolute magnitude) at which a galaxy could have been detected at its redshift, $L_\mathrm{faint}$ (or $M_\mathrm{faint}$). 
In practice, if the sample is limited by the observed magnitude, $m$, down to the limiting depth $m \leq m_{\mathrm{lim}}$, then 
\begin{equation}
 \log(L_\mathrm{faint}) = \log(L) + 0.4 \ (m - m_{\mathrm{lim}})
\end{equation}
which, in terms of absolute magnitude, gives us
\begin{equation}
\label{eq_Mabs_lim}
M_\mathrm{faint} = M - (m - m_{\mathrm{lim}}) ~.
\end{equation}
In each redshift bin, we conservatively considered the 20\% highest redshift galaxies (i.e., those that are closest to the upper limit of the redshift bin).  The corresponding absolute magnitude completeness limit, $M_\mathrm{lim}$, was then defined by the absolute magnitude for which 90\% of that upper-limit population had an absolute magnitude $M < M_\mathrm{faint}$.

Given that our detection images combine all the CLAUDS and HSC-SSP passbands (i.e., $u,u^*,g,r,i,z,y$), every band contributes to the completeness limit. 
Assuming that a source is detected as long as it is bright enough in at least one of the bands, we derived the effective absolute magnitude completeness limit of our sample, $M_\mathrm{lim}$, as the faintest absolute magnitude completeness limit computed in all the bands, i.e., 
 \begin{equation}
 \label{eq_Mabs_lim_eff}
M_\mathrm{lim} = \max_b (M^b_\mathrm{lim}), ~~ \mathrm{for} \ b=u,u^*,g,r,i,z,y, \,
\end{equation}
where $M^b_\mathrm{lim}$ is the absolute magnitude completeness limit derived from the limiting depth of the passband $b$, following Equation~\ref{eq_Mabs_lim}.

As detailed in Sect. \ref{sect_data}, our survey contains two layers of different depths: Deep and Ultra-Deep.
The advantage of such structure was twofold. 
1) The different depths of the Deep and Ultra-Deep layers allowed us to fine-tune our method of measuring the completeness limit by cross-matching the results from the two layers; with this, we ensure that we did not miss more than 10~percent of galaxies in the faintest magnitude bin. 
2) In order to take the best advantage of our survey, we adopted a wedding cake approach where the bright end of the LF comes from the Deep layer, down to the corresponding completeness limit, below which the very faint end relies on the Ultra-Deep layer.

\subsubsection{LF measurement}

Given the depth of the two layers of our survey, we decided to adopt highly conservative absolute magnitude completeness limits, as discussed in the previous section, which allowed us to measure the FUV, NUV and U-band LFs without incompleteness correction at $M < M_{lim}$. 

However, aiming to validate our method, we also measured the LFs with the tool ALF \citep{Ilbert2005}, using two different LF estimators: the $V_{max}$ \citep{Schmidt1968} and SWML \citep[the step-wise maximum likelihood;][]{Efstathiou1988}. We verified that these two estimators were in good agreement with our uncorrected estimation of the LF down to our adopted completeness limit, which de facto confirmed our estimation of the completeness limit \citep[$V_{max}$ and SWML estimators are known to diverge below the completeness limit;][]{Ilbert2005, Moutard2016b}.

\begin{table*}
\begin{center}
\caption{Demographics of our survey.  Numbers in brackets indicate objects that were not used in constructing the LFs. Objects in the Overlap column were present in both the Deep and Ultra-Deep layers and were counted only once in building the LFs. \label{table_demo}}
\begin{tabular}{l*{5}{c}}
\multicolumn{5}{c}{Number of galaxies} \\
\hline \\[-5mm]
\hline \\ [-3mm]
Redshift bin & Deep $^{(a)}$ & Ultra-Deep $^{(b)}$ & Overlap $^{(c)}$ & Total used $^{(d)}$ \\[0mm] 
\hline \\[-3mm]
$0.05 < z < 0.3$ & 201,617 & (\textit{18,073}) & ----- & 201,617 \\[1mm] 
$0.3 < z < 0.45$ & 296,940 & (\textit{30,598}) & ----- & 296,940 \\[1mm] 
$0.45 < z < 0.6$ & 331,144 & (\textit{31,291}) & ----- & 331,144 \\[1mm] 
$0.6 < z < 0.9$ & 735,345 & 84,059 &  63,473 & 755,931\\[1mm] 
$0.9 < z < 1.3$ & 1,142,045 & 143,771 & 93,381 & 1,192,435\\[1mm] 
$1.3 < z < 1.8$ & 830,481 & 134,570 & 68,037 & 897,014\\[1mm] 
$1.8 < z < 2.5$ & 566,007 & 102,226 & 58,784 & 609,449\\[1mm] 
$2.5 < z < 3.5$ & (\textit{277,050}) & 54,977 & ----- & 54,977  \\[1mm] 
\hline \\[-3mm] 
  & (\textit{277,050}) & (\textit{79,962}) &  &  \\[1mm]
$0.05 < z < 3.5$ & 4,103,579 & 519,603 & 283,675 & 4,339,507 \\
\hline \\[-5mm] 
\hline \\[-3mm]
\multicolumn{5}{l}{\begin{footnotesize} $^{(a)}$ Number of galaxies in the Deep layer (cf. Equation~\ref{eq_deep_sel} and Fig. \ref{fig_footprint}). \end{footnotesize}} \\
\multicolumn{5}{l}{\begin{footnotesize} $^{(b)}$  Number of galaxies in the Ultra-Deep layer (cf. Equation~\ref{eq_udeep_sel} and Fig. \ref{fig_footprint}). \end{footnotesize}} \\
\multicolumn{5}{l}{\begin{footnotesize} $^{(c)}$ Number of galaxies present in both the Deep and Ultra-Deep layers. \end{footnotesize}} \\
\multicolumn{5}{l}{\begin{footnotesize} $^{(d)}$  Total number of galaxies considered in the present study: (d)=(a)+(b)-(c). \end{footnotesize}} \\

\end{tabular}
\end{center}
\end{table*}

By definition, the luminosity function, $\phi(L)dL$, is defined as the comoving number density of galaxies with luminosity between $L$ and $L+dL$, or in term of absolute magnitude $M$, $\phi(M) dM =\phi(L)d(-L)$.
To compute the FUV, NUV and U-band luminosity functions, we first selected a sample of 4,380,629 galaxies with $z < 3.5$ in the Deep layer, which covers an effective area (i.e., after masking) of 17.02 deg$^2$ down to
\begin{eqnarray}
(U\le26.9) \ \cup \ (g\le26.3) \ \cup\  (r\le25.9) \ \cup \nonumber\\
\ (i\le25.5) \ \cup \ (z\le24.9) \ \cup \ (y\le24.0) 
\label{eq_deep_sel}
\end{eqnarray} 
and 599,565 galaxies with $z < 3.5$ in the Ultra-Deep layer, which covers an effective area of 1.45 deg$^2$ down to
\begin{eqnarray}
(U\le27.4) \ \cup \ (g\le26.9) \ \cup\  (r\le26.7) \ \cup \nonumber\\
\ (i\le26.4) \ \cup \ (z\le25.9) \ \cup \ (y\le24.8) ~.
\label{eq_udeep_sel}
\end{eqnarray} 
As discussed in Sect. \ref{sect_phys_param}, we restricted our analysis to the redshift ranges $0.05\le z \le 3.5$ in UV and $0.05\le z \le 2.5$ in U-band, where both photometric redshifts and absolute magnitudes are well constrained. We defined eight contiguous redshift bins which were chosen by considering the observed bands used to derive the absolute magnitudes: 
$0.05 < z \leq 0.3$, $0.3 < z \leq 0.45$, $0.45 < z \leq 0.6$, $0.6 < z \leq 0.9$, $0.9 < z \leq 1.3$, $1.3 < z \leq 1.8$, $1.8 < z \leq 2.5$, and $2.5 < z \leq 3.5$. 

Table \ref{table_demo} summarises the corresponding numbers of galaxies available in the two layers of our survey, as well as the numbers of galaxies we finally considered to measure the LFs after combining the two layers. 
Note that we only used the Deep layer to measure the LFs at $0.05 < z \leq 0.6$, given the cosmic variance affecting the Ultra-Deep layer at low redshift due to the limited volume it probes.
On the other hand, concerning the last redshift bins we considered for the UV and U-band LFs (namely, $2.5 < z \leq 3.5$ and $1.8 < z \leq 2.5$, respectively), we only used the Ultra-Deep layer, given the limited depth of our $g,r,i,z,y$ data in the Deep layer. In total, we thereby made use of 4,339,507 galaxies to measure the FUV, NUV and U-band LFs 
%

\begin{figure*}
\center
\includegraphics[width=\hsize, trim = 6cm 6cm 7cm 6cm, clip]{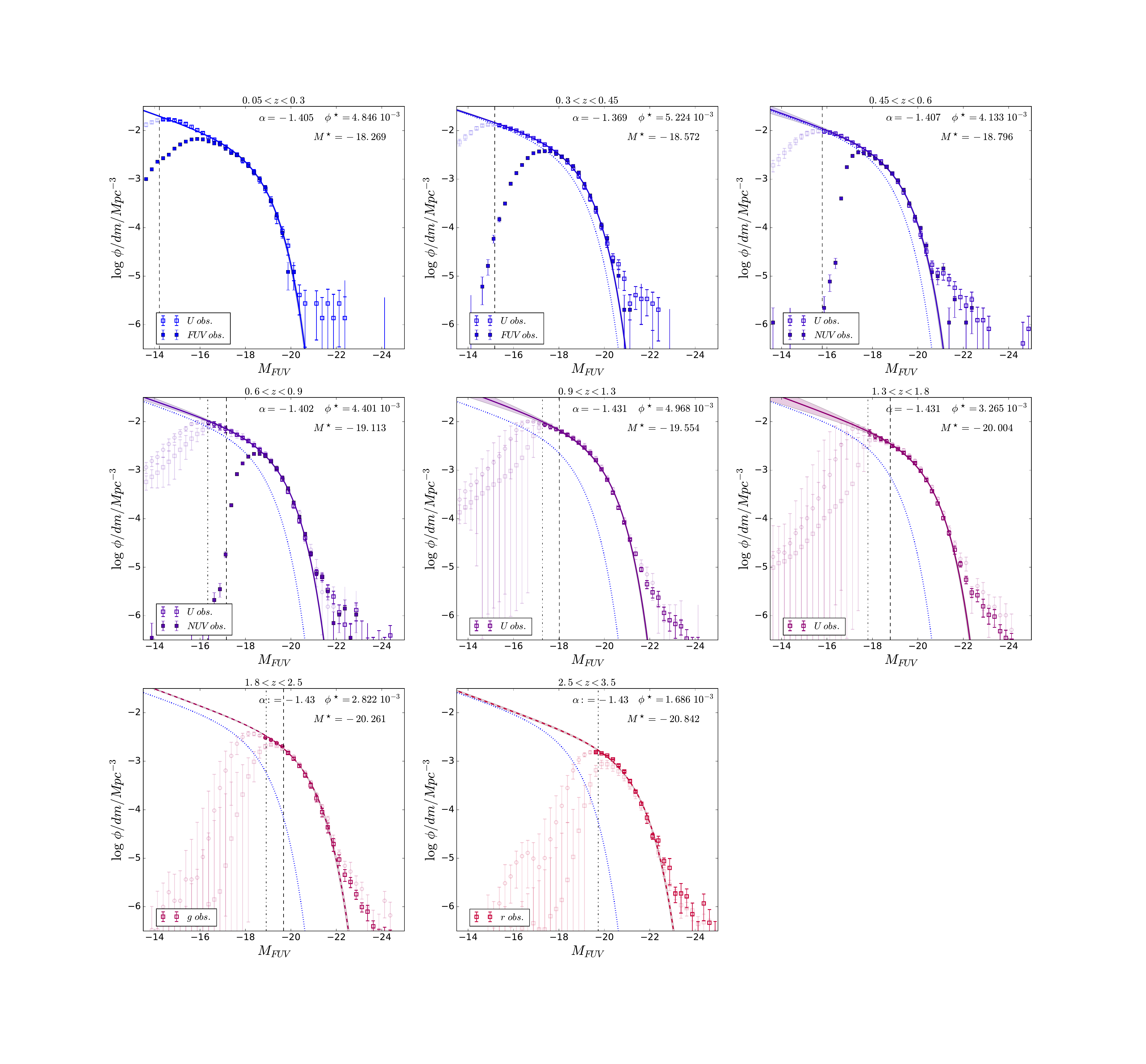}
\caption{FUV luminosity function measured at redshift $0.05 < z \leq 0.3$, $0.3 < z \leq 0.45$, $0.45 < z \leq 0.6$, $0.6 < z \leq 0.9$, $0.9 < z \leq 1.3$, $1.3 < z \leq 1.8$, $1.8 < z \leq 2.5$, and $2.5 < z \leq 3.5$. 
At $z < 0.9$, the luminosity function based on rest-frame FUV extrapolated from our U-band data is compared with the luminosity function derived from direct observation of the FUV luminosity in the GALEX FUV and NUV passbands (filed squares). 
At $0.6 < z < 2.5$, the faint end of the LF comes from our Ultra-Deep layer (open circles). At $2.5 < z < 3.5$ the whole LF is based on the Ultra-Deep layer. 
In each panel, the (solid and dashed) curve and associated envelop show our best Schechter fit and corresponding 1$\sigma$ uncertainty. At $1.8 < z < 3.5$, the slope was set to $\alpha=-1.43$ (see text) and the best Schechter fits are plotted with dashed curves.
Vertical dashed and dash-dotted lines show, respectively, the location of the Deep and Ultra-Deep layer completeness limits (see Table \ref{table_param}).
\label{fig_FUV_LFs}  }
\end{figure*}

\begin{figure*}
\center
\includegraphics[width=\hsize, trim = 6cm 6cm 7cm 6cm, clip]{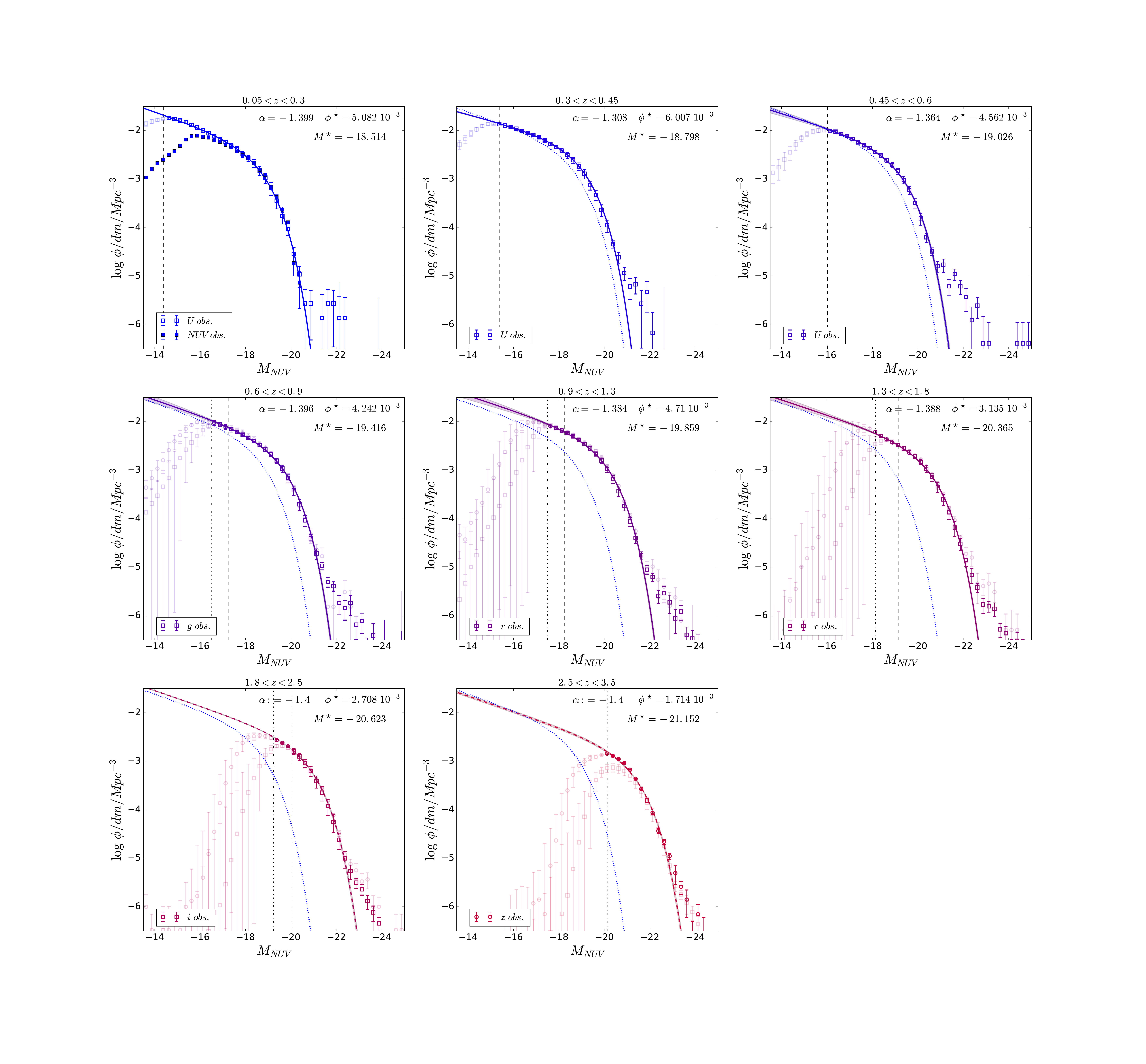}
\caption{NUV luminosity function measured at redshift $0.05 < z \leq 0.3$, $0.3 < z \leq 0.45$, $0.45 < z \leq 0.6$, $0.6 < z \leq 0.9$, $0.9 < z \leq 1.3$, $1.3 < z \leq 1.8$, $1.8 < z \leq 2.5$, and $2.5 < z \leq 3.5$. 
At $z < 0.3$, the luminosity function based on rest-frame NUV extrapolated from our U-band data is compared with the luminosity function derived from direct observation of the NUV luminosity in the GALEX NUV passband (filed squares). 
Similarly to Fig. \ref{fig_FUV_LFs}, at $0.6 < z < 2.5$, the faint end of the LF comes from our Ultra-Deep layer (open circles), and at $2.5 < z < 3.5$ the whole LF is based on the Ultra-Deep layer.
In each panel, the (solid and dashed) curve and associated envelop show our best Schechter fit and corresponding 1$\sigma$ uncertainty. At $1.8 < z < 3.5$, the slope was set to $\alpha=-1.4$ (see text) and the best Schechter fits are plotted with dashed curves.
Vertical dashed and dash-dotted lines show, respectively, the location of the Deep and Ultra-Deep layer completeness limits (see Table \ref{table_param}).
\label{fig_NUV_LFs}  }
\end{figure*}

\begin{figure*}
\center
\includegraphics[width=\hsize, trim =6cm 6cm 7cm 6cm, clip]{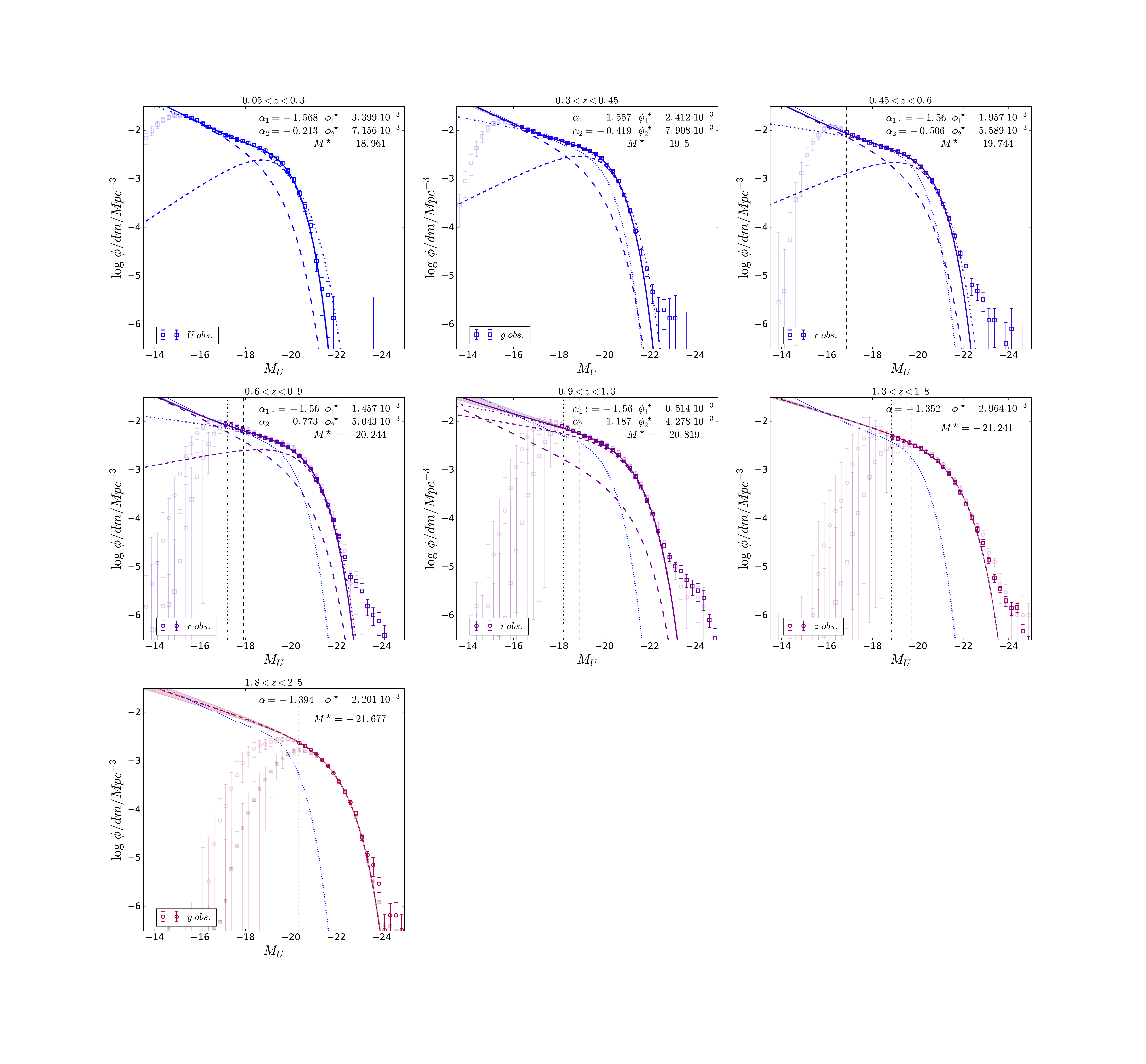}
\caption{U-band luminosity function measured at redshift $0.05 < z \leq 0.3$, $0.3 < z \leq 0.45$, $0.45 < z \leq 0.6$, $0.6 < z \leq 0.9$, $0.9 < z \leq 1.3$, $1.3 < z \leq 1.8$, and $1.8 < z \leq 2.5$. 
Similarly to Figs. \ref{fig_FUV_LFs} and \ref{fig_NUV_LFs}, at $0.6 < z < 1.8$, the faint end of the LF comes from our Ultra-Deep layer (open circles), and at $1.8 < z < 2.5$ the whole LF is based on the Ultra-Deep layer.
In each panel, the solid line and associated envelop show our best Schechter fit and corresponding 1$\sigma$ uncertainty. At $z < 1.3$, both single Schechter (dash-dotted lines) and double Schechter (solid lines and associated envelops) functions are shown (the two components of the double Schechter function are shown with dashed lines). At $0.45 < z < 1.3$, the faint-end slope of the double Schechter function was set to $\alpha_1=-1.56$. At $1.3 < z < 2.5$, only a single Schechter was considered (see text).
Vertical dashed and dash-dotted lines show, respectively, the location of the Deep and Ultra-Deep layer completeness limits (see Table \ref{table_param}).
\label{fig_U_LFs}  }
\end{figure*}

Figure \ref{fig_FUV_LFs} shows the FUV LF we measured in the eight redshift bins  we defined from $z = 0.05$ to $z = 3.5$. For each redshift bin, we specified the observed passband in which the FUV absolute magnitude was generally derived. At lower redshifts, our CLAUDS+HSC-SSP measurements involve an extrapolation blueward of the observed U-band, and we verify that this extrapolation is reasonable using GALEX data as follows. In the four lowest redshift bins, we compare the LF measured from observed GALEX FUV and NUV (which minimizes the k-correction) with the LF measured from CLAUDS U-band observations. As one can see, the two LF measurements are in very good agreement down to $M_\mathrm{FUV} \simeq -17, -18, -18$ and $-19$ at $0.05 < z \leq 0.3$, $0.3 < z \leq 0.45$, $0.45 < z \leq 0.6$, $0.6 < z \leq 0.9$, respectively, where we reach the depth of the GALEX observations.\footnote{Note that GALEX fluxes measured with EMphot only use u band priors down to \oldu$\sim$25 (see Sect. \ref{sect_data}).} This agreement suggests that the FUV absolute magnitude we derived from extrapolation of U-band observations is reliable.
Similarly, Fig. \ref{fig_NUV_LFs} shows the NUV LF in the same redshift bins. In the lowest redshift bin, we compare the LF measured from GALEX NUV and from CLAUDS U-band, and one can see that both LF measurements are in very good agreement down to $M_\mathrm{NUV} \simeq -17$, the depth of the GALEX observations. As with the FUV measurements, this agreement suggests that the NUV absolute magnitude we derived from the extrapolation of U-band observations is well constrained. Finally, in Fig. \ref{fig_U_LFs}, we show the U-band LF we measured in the seven redshift bins from $z = 0.05$ to $z = 2.5$

In Figs. \ref{fig_FUV_LFs}, \ref{fig_NUV_LFs} and \ref{fig_U_LFs} we showed the LFs we measured in the Deep (squares) and Ultra-Deep (circles) layers. 
We adopted the wedding cake approach presented in the previous section when the comoving volume of the redshift bin was large enough to be characterized by an average density close to that of the Universe at that redshift (i.e., when the faint end of the LF is not dominated by the so-called cosmic variance that we discuss in the next section). 
One can see how at $z > 0.6$, the faint end of the LF is based on the Ultra-Deep layer down to the associated completeness limit, while the rest of the LF is derived from the Deep layer.  
On the other hand, our LF measurements in the last redshift bins we considered for the UV ($2.5 < z \leq 3.5$) and U-band ($1.8 < z \leq 2.5$) were entirely based on the Ultra-Deep layer, given the very small contribution of the Deep layer at those redshifts (because of its fairly bright $g,r,i,z,y$ limits).

\subsubsection{LF uncertainties}

In addition to the Poissonian error ($\sigma_{Poi}$) usually taken into account,  LF measurements suffers from two addtional main sources of uncertainty: the error on the luminosity or absolute magnitude ($\sigma_M$), as described in Sect. \ref{sect_absmag_err}, and the so-called cosmic variance $(\sigma_{cv})$, which is due to large-scale inhomogeneities in the spatial distribution of galaxies in the Universe.

These additional sources of uncertainty can have an important contribution to the total error budget and therefore need to be accounted for. For instance,  cosmic variance has been shown to represent a fractional error of $\sigma_{cv} = 10-15\%$ for massive ($M_*\ge 10^{11} M_{\odot}$)  galaxies with number densities of $\phi < 10^{-3}$ Mpc$^{-3}$ in a 2-deg$^2$ survey, against $\sigma_{cv} \sim 6-8\%$ in a 20-deg$^2$ survey. At the same time, the cosmic variance contribution to the error budget is small compared to the Poissonian error for very massive --i.e., rare-- galaxies, while it dominates the error budget for lower mass --i.e., more abundant-- galaxies \citep[see][for a discussion of these issues]{Moutard2016b}.  We may expect a similar effect on the luminosity function, where the very bright end suffers from large cosmic variance and suffers from an even larger Poissonian error, while at fainter magnitudes a modest cosmic variance dominates a very small Poissonian error.

Aiming to estimate the contribution of the cosmic variance affecting our LF measurements, we adopt the procedure followed by \citet{Moutard2016b}, which is based on a method introduced by \citet{Coupon2015}. In brief, at given area $a$, we derived cosmic variance from Jackknife resampling of N patches with area $a$, for patch areas ranging from $a=0.2$ to $1.6$ deg$^2$. The cosmic variance measured using subareas of our survey is then extrapolated to the total area, namely, $a=18.29$ deg$^2$ and $a=1.54$ deg$^2$ in the Deep and Ultra-Deep layers, respectively \citep[for more details on the method, please refer to][]{Coupon2015, Moutard2016b}.

The last source of uncertainty that we need to consider comes from the error on the absolute magnitude, $\sigma_{M}$, as defined in Sect. \ref{sect_absmag_err}. To convert $\sigma_{M}$ into an error on the number density, $\sigma_{\phi, M}$, we generated 200 mock catalogues with absolute magnitudes perturbed according to $\sigma_{M}$ (cf. Equation~\ref{eq_err_MABS}) and measured the 1$\sigma$ dispersion of the perturbed LFs.  

The total uncertainty, $\sigma_{\phi}$, affecting the luminosity function at each magnitude bin is then calculated by combining the three sources of error in quadrature,
\begin{equation}
\sigma_{\phi} = \sqrt{ ~\sigma_{Poi}^2 + \sigma_{cv}^2~ +  ~\sigma_\mathit{\phi, M}^2} ~,
\label{eq_err_LF}
\end{equation}
and plotted in Figs. \ref{fig_FUV_LFs},\ref{fig_NUV_LFs} and \ref{fig_U_LFs}.

Note that although $\sigma_{\phi, M}$ is a good estimation of the contribution of the absolute magnitude error in the LF error budget, it cannot take into account the so-called Eddington bias, whose effects we treat as discussed in Sect.~\ref{sect_fit_Edd_treat}.

\begin{figure*}
\center
\includegraphics[width=0.33\hsize, trim = 0cm 0cm 1.5cm 1cm, clip]{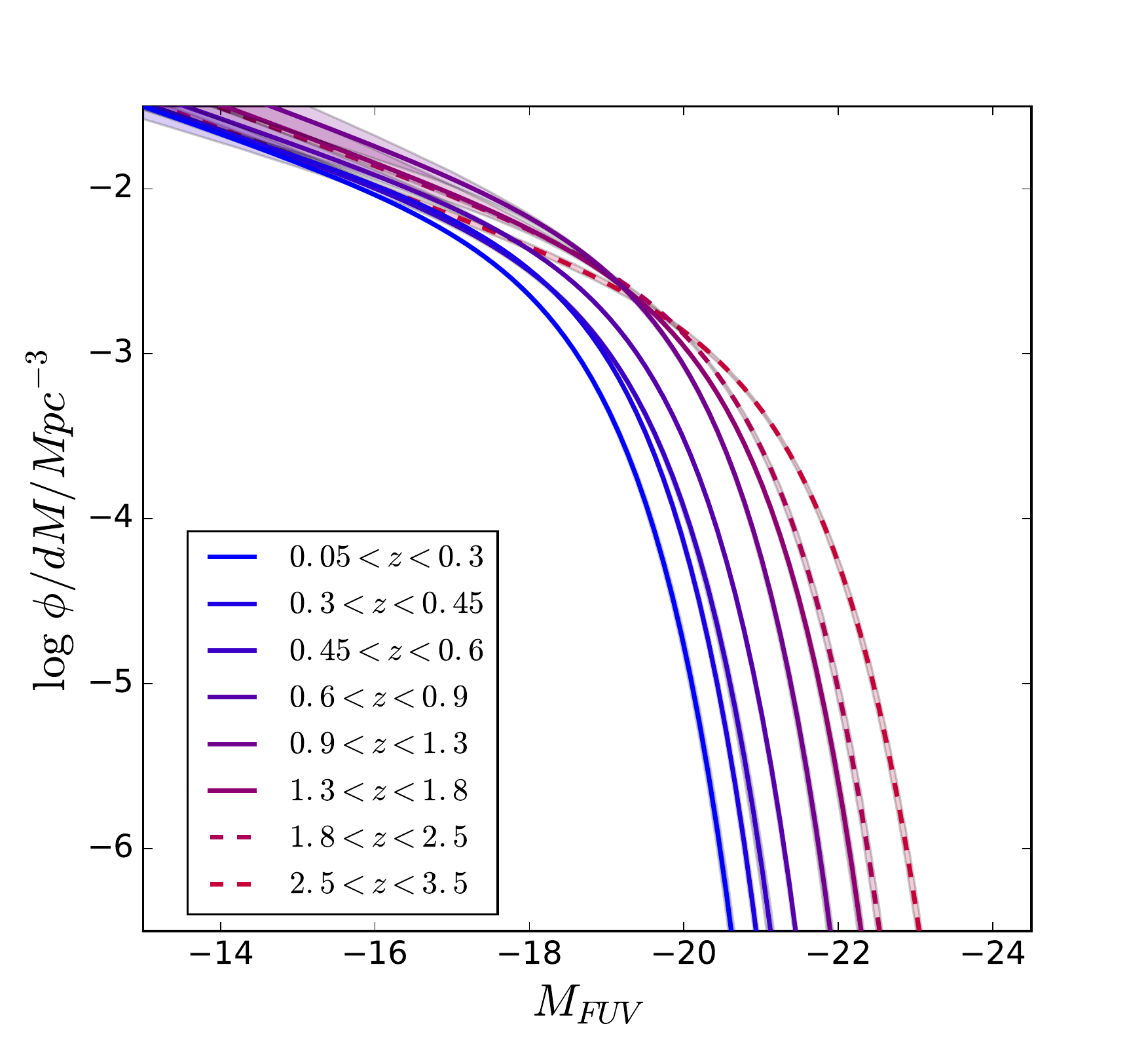}
\includegraphics[width=0.33\hsize, trim = 0cm 0cm 1.5cm 1cm, clip]{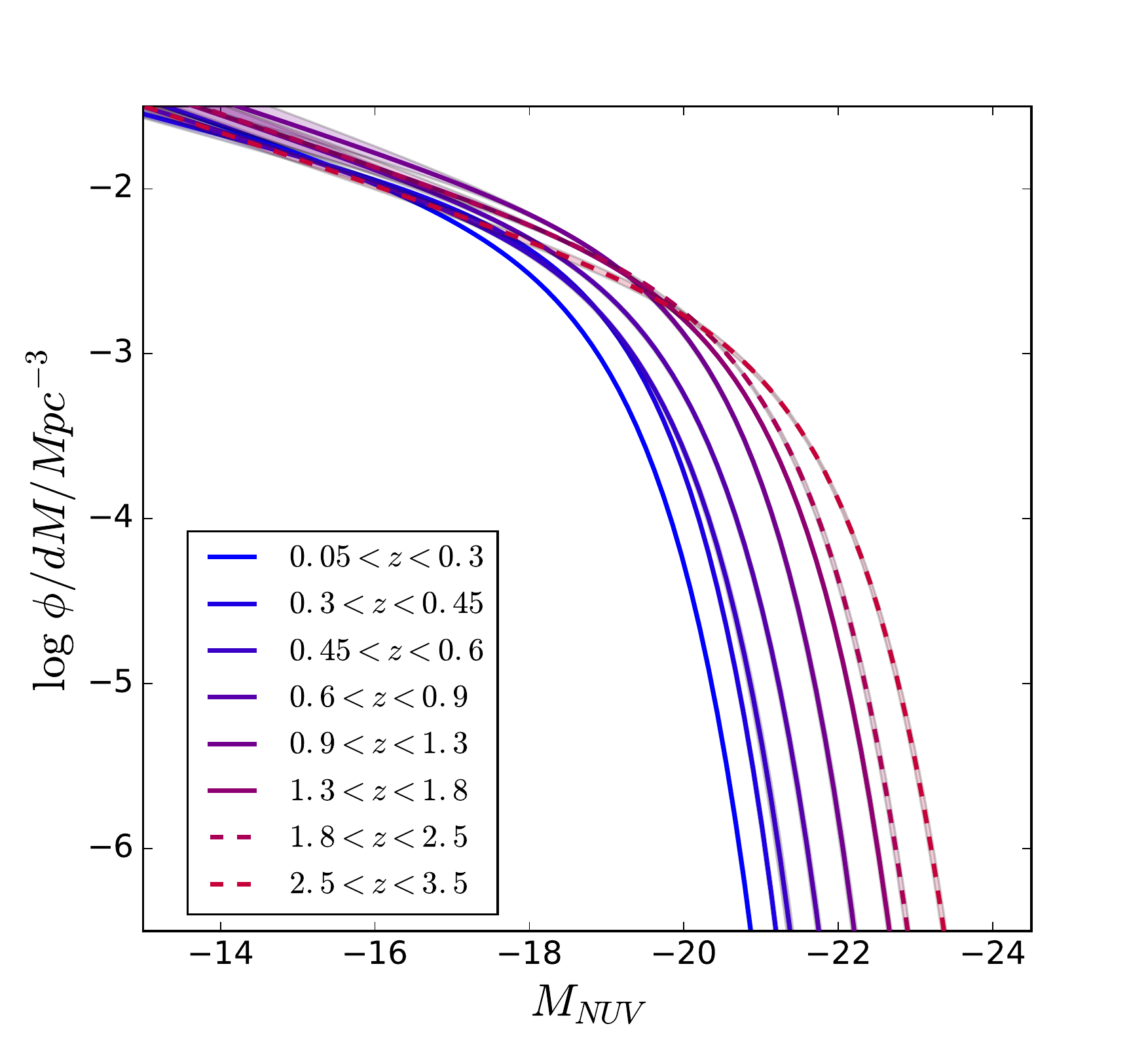}
\includegraphics[width=0.33\hsize, trim = 0cm 0cm 1.5cm 1cm, clip]{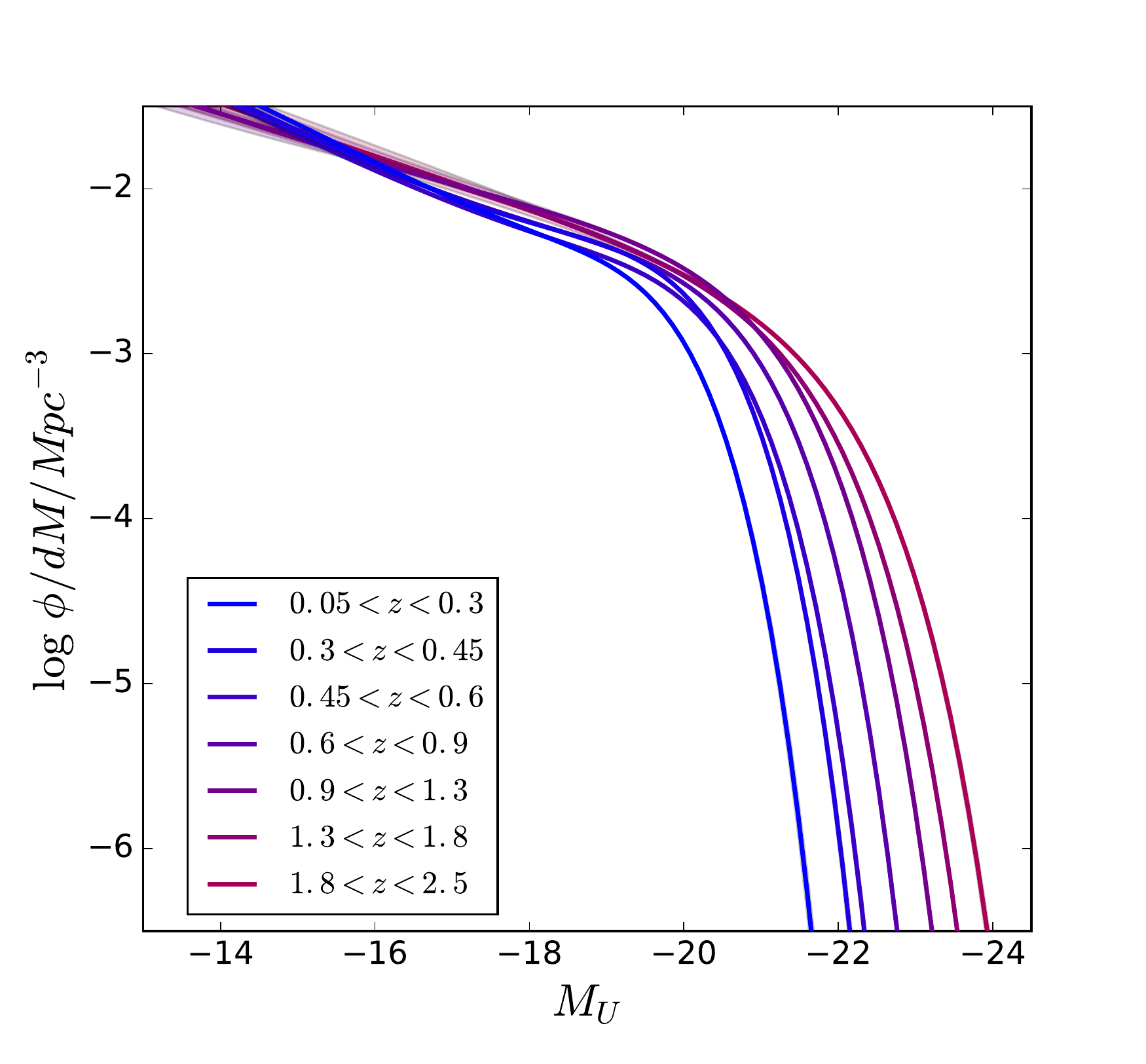}
\caption{Redshift evolution of the FUV, NUV and U-band luminosity functions. Only the LF Schechter-based parametric form and corresponding 1$\sigma$ uncertainty are shown here, with (solid and dashed) curves and associated envelopes, respectively. Dashed curves show Schechter-based fits with slopes that were not free to vary (see text).
\label{fig_LF_evol}  }
\end{figure*}

\begin{figure*}
\center 
\includegraphics[width=0.33\hsize, trim = 0.3cm 0cm 0.4cm 0cm, clip]{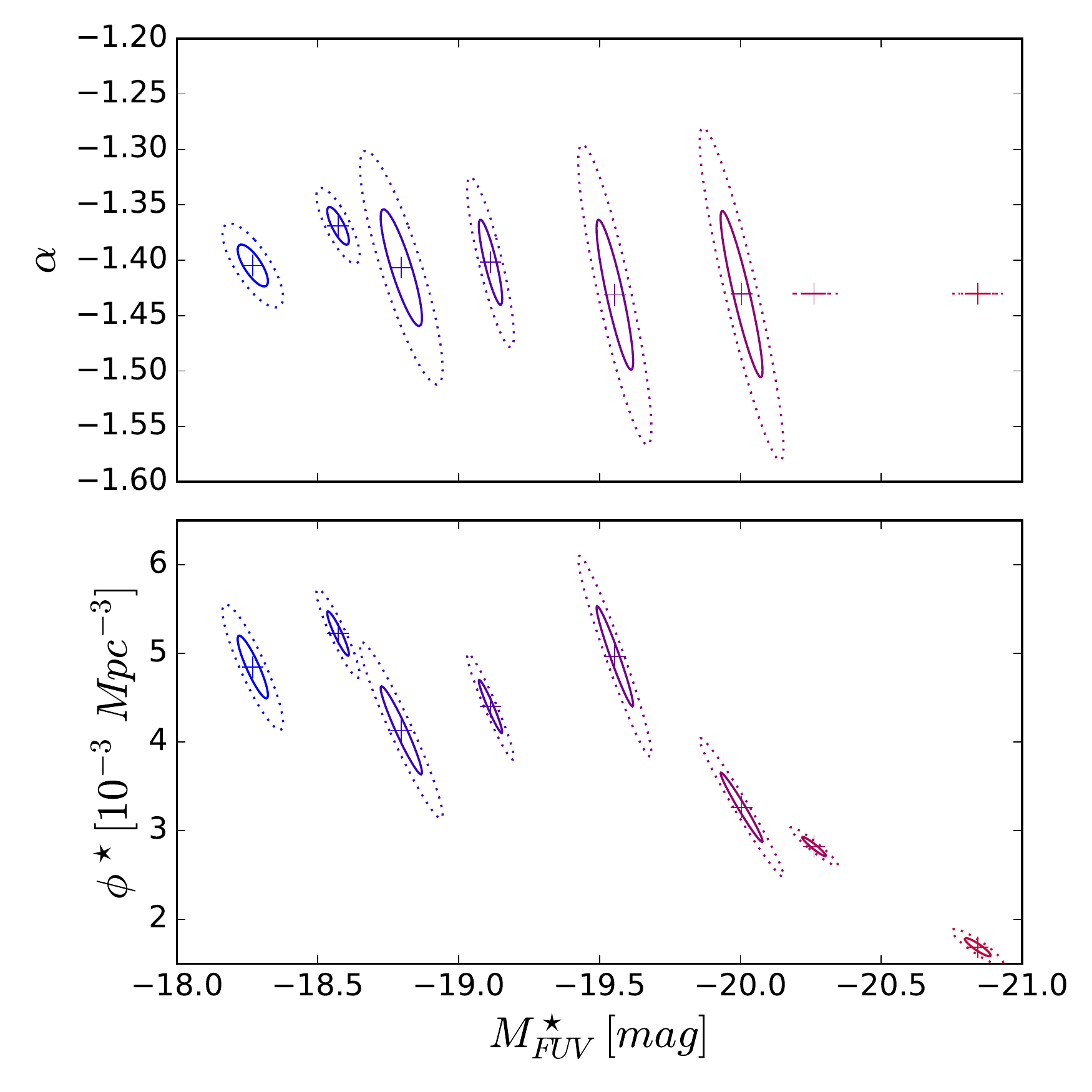}
\includegraphics[width=0.33\hsize, trim = 0.55cm 0cm 0.15cm 0cm, clip]{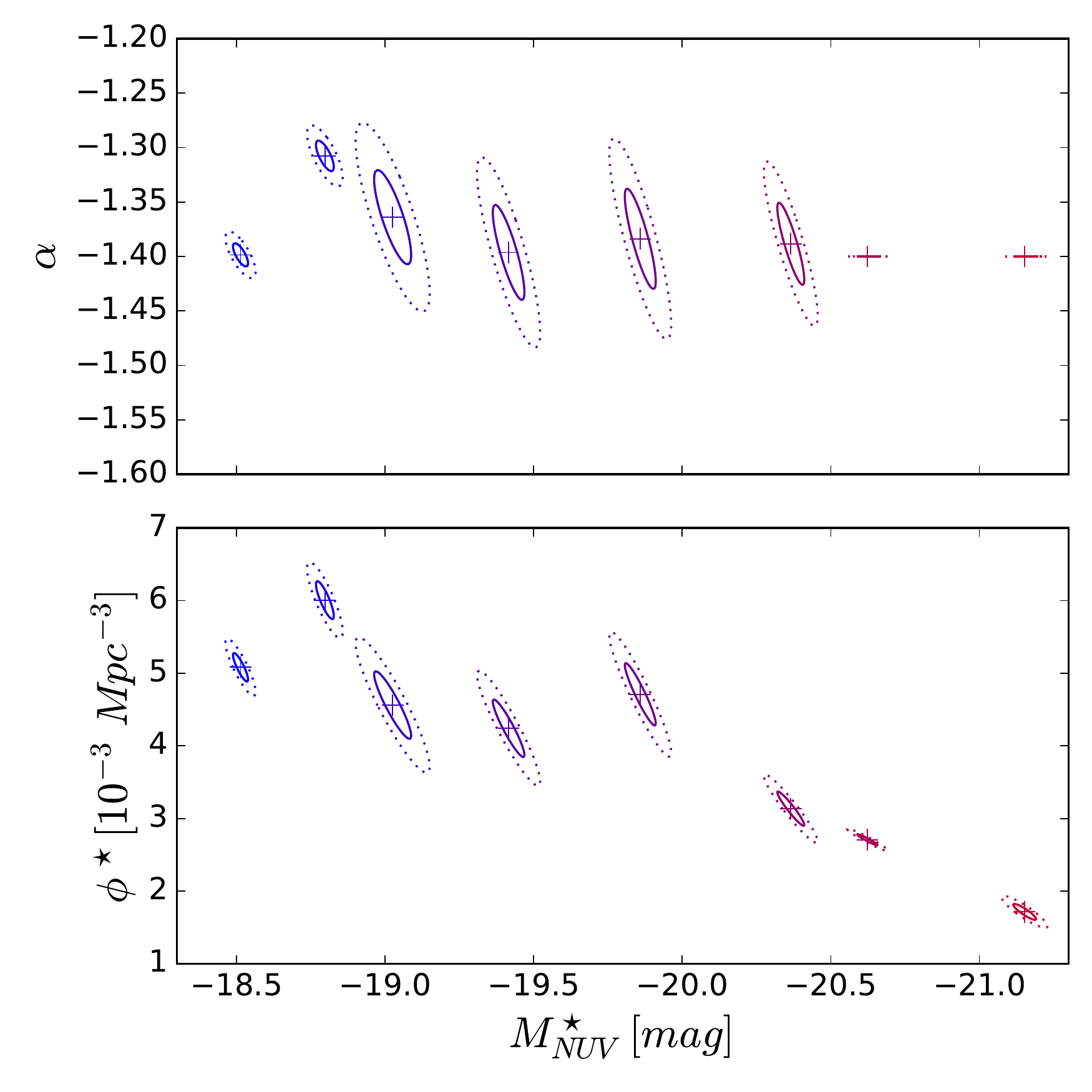}
\includegraphics[width=0.33\hsize, trim = 0.3cm 0cm 0.4cm 0cm, clip]{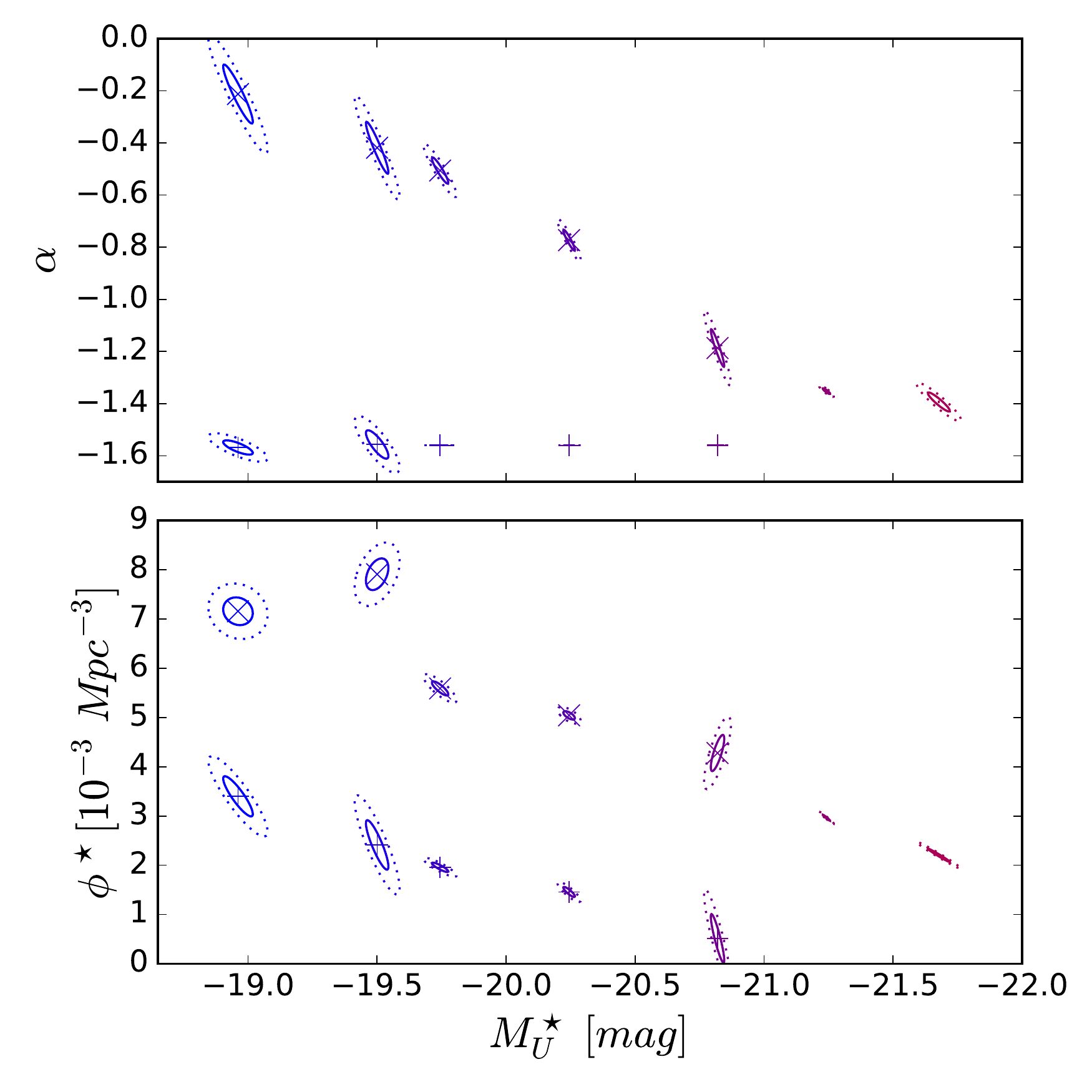}
\caption{FUV, NUV and U-band LF best-fitting Schechter parameters $\alpha$ and $\phi^\star$ as a function of $M^\star$ (crosses), and associated 1$\sigma$ and 2$\sigma$ confidence regions (solid and dotted ellipses, respectively). Consistently with the fitting procedure described in Sect. \ref{sect_fit_forms}, the U-band LF Schechter parameters shown in the right-hand panel are those of a single Schechter function at $1.3 < z < 2.5$, and a double Schechter function at $0.05 < z < 1.3$: the faint-end component parameters ($\alpha_1$, $\phi^\star_1$) and bright-end component parameters ($\alpha_2$, $\phi^\star_2$) are shown with "$+$" and "$\times$", respectively.
\label{fig_LFparam_evol}  }
\end{figure*}

\subsection{Redshift evolution of the luminosity functions}
\label{sect_LFmeasurements}

\subsubsection{Fitting method and Eddington bias treatment}
\label{sect_fit_Edd_treat}

Eddington bias \citep{Eddington1913}  affects the observed slope of the bright end of the luminosity function by converting the statistical error on the luminosities of the more abundant (usually fainter) galaxies into a systematic boost of the number of less abundant  (usually brighter) galaxies.  The result of this effect is that the observed slope of the luminosity  function is shallower than the underlying reality.  The same effect affects the observed stellar mass functions, where the dominant effect is that of the scattering of lower-mass galaxies into the higher-mass population. 

Several authors have addressed the Eddington bias over the past few years, especially regarding the high-mass end of the stellar mass function (SMF).  These studies have implicated the effect as biasing the rather mild evolution of the high-mass end of the SMF at $z < 1$  \citep[e.g.,][]{Matsuoka2010, Ilbert2013, Moutard2016b}; at $z>4$, where the SMF evolution is stronger, stellar mass uncertainties are very large and so still have to be taken into account \citep[e.g.,][]{Caputi2011, Grazian2015, Davidzon2017}. 
Although less discussed in the literature, the LF bright end measurements suffer from a similar effect that needs to be accounted for. 

In the present study, we accounted for the effects of Eddington bias by following a procedure similar to that described in \citet{Ilbert2013}, as we fitted our LF measurement through $\chi^2$ minimization.
In this, we only consider the statistical uncertainties (Poisson and cosmic variance) in the $\chi^2$ calculation during the fitting process, while the absolute magnitude uncertainty $\sigma_M$ is taken into account through convolution with the fitted Schechter parametric form(s), which is thereby corrected for the Eddington bias.  Adapting the approach of \citet{Moutard2016b}, we consider an estimate of $\sigma_M$ that varies with absolute magnitude and redshift, $\sigma_M(M, z)$, in order to avoid over-correction of the Eddington bias.

Finally, one may notice from Figs. \ref{fig_FUV_LFs}, \ref{fig_NUV_LFs} and \ref{fig_U_LFs} that the extremely bright ends of our LF measurements, typically for comoving densities $< 10^{-5}$ Mpc$^{-3}$, suffer from uncertainties that are significantly larger than what one could expect from purely Poissonian errors. 
As discussed in Appendixes \ref{app_LF_perfield} and \ref{app_LF_sourcetype}, this is most probably due to the contamination by stars and QSOs, the identification and cleaning of which depends on the depth of our observations that varies across the survey. A very small number of interlopers is indeed sufficient to affect the actual number of extremely bright and rare galaxies. 
In any event, we verified that this contamination of the extremely-bright ends had not a significant impact on the fitting of the LFs that is discussed in the following.\footnote{In practice, we found that the difference between the best-fitting parameters obtained by considering or excluding the LF points with comoving densities $< 10^{-5}$ Mpc$^{-3}$ was smaller than the typical error on those parameters.}

\subsubsection{FUV, NUV \& U-band LF fitting}
\label{sect_fit_forms}

As can be seen in Figs. \ref{fig_FUV_LFs} and \ref{fig_NUV_LFs}, the Schechter parametric form \citep{Schechter1976} appears to be well suited to the fitting of the FUV and NUV LFs down to the completeness limits of our survey and, at least, between $z = 0.05$ and $z = 3.5$.

We therefore fitted the FUV and NUV LFs with the classical Schechter function defined as 
\begin{equation}
\phi(L) \ dL = e^{-\frac{L}{{L}^\star}} \ \phi^\star \left(\frac{L}{{L}^\star} \right) ^{\alpha}  \frac{dL}{{L}^\star} ~,
\label{eq_Sch}
\end{equation}
which, in term of absolute magnitude, can be written as
\begin{equation}
\phi(M) \ dM = \frac{\ln 10}{2.5} \ \phi^{\star}  \left( 10^{ 0.4 \Delta M} \right)^{ \alpha + 1}  \exp \left( -10^{ 0.4 \Delta M } \right) \ dM ~,
\label{eq_singleSch_mag}
\end{equation}
with $\Delta M = M^\star - M$.

While all three Schechter parameters  are well constrained at $0.05 < z \leq 1.8$, the slope $\alpha$ and normalisation $\phi^\star$ start being poorly constrained at $1.8 < z \leq 2.5$ and are no longer constrained at $z > 2.5$.
One of the strengths of our dataset is its ability to probe the bright end of the LF, thanks to the large area covered, the location of which is well traced by the characteristic absolute magnitude $M^{\star}$.  Given the  stability of $\alpha$ at $0.05 < z \leq 1.8$, we can help constraint  $M^{\star}$ at $z > 1.8$ by setting $\alpha$ at $z>1.8$ to its average value at $z \leq 1.8$:  $\alpha_{z > 1.8} = \mathrm{const.} = \overline{\alpha}_{z<1.8}$. 
Our Schechter functional fits are plotted in Figs.~\ref{fig_FUV_LFs} and \ref{fig_NUV_LFs} with solid lines at $z < 1.8$ and dashed lines at $z > 1.8$, and the values of the Schechter parameters are listed in Table \ref{table_param}.

While the FUV and NUV LFs are well described by the Schechter function (Eq.~\ref{eq_singleSch_mag}), this is not the case for the U-band LF (Fig. \ref{fig_U_LFs}). Here, the LF shape deviates from the classical Schechter form at the faint end, where a clear upturn can be seen around $M_U \sim -17$, at least at low redshift (where our completeness limit is the faintest). 
Where needed, we therefore adopt a double-Schechter to fit the U-band LF, as defined by
\begin{equation}
\phi(L) \ dL = e^{-\frac{L}{{L}^\star}} \ \left[ \phi^\star_1 \left(\frac{L}{L^\star} \right) ^{\alpha_1} + \phi^\star_2 \left(\frac{L}{L^\star} \right) ^{\alpha_2} \right] \ \frac{dL}{L^\star} ~,
\label{eq_doubleSch}
\end{equation}
or, in term of absolute magnitude, 
\begin{eqnarray}
\phi(M) \ dM = \frac{\ln 10}{2.5} ~
\left[ \ \phi_1^{\star}  \left( 10^{ 0.4 \Delta M  } \right)^{ \alpha_1 + 1} 
 + ~ \phi_2^{\star}  \left( 10^{ 0.4 \Delta M } \right)^{ \alpha_2 + 1} \ \right] \nonumber\\ 
~~ \times~~ \exp \left( -10^{ 0.4 \Delta M } \right) \ dM ~~ 
\label{eq_doubleSch_mag}
\end{eqnarray}
with $\Delta M = M^\star - M$ and, in our case, $\alpha_1 < \alpha_2$.
For further detail about the relevance of fitting the U-band LF with a double-Schechter function, please refer to Appendix \ref{app_U_LF_fitting}.

While all the double-Schechter parameters were well constrained at $z \leq 0.5$ and could be fitted simultaneously, we had to constrain the parameters of the faint end before fitting the LF at higher redshift (notably due to the difference between the uncertainties affecting the Deep and Ultra-Deep LF contributions, which tend to drastically reduce the Ultra-Deep layer contribution at the faint end of the LF in the $\chi^2$ fitting). 
To constrain the fitting at $z > 0.5$, we set the faint-end slope $\alpha_1$ to the its average value at $z \leq 0.45$:  $\alpha_1 (z > 0.45) = \mathrm{const.} = \overline{\alpha_1}_{z<0.45}$.
On the other hand, at $1.3 < z \leq 2.5$, the completeness limit prevented us from observing a second Schechter component at the faint end of the LF. At these high redshifts we only fitted the LF with a single Schechter function.
Finally, we also showed the parametric form we would obtain by fitting the U-band LF with a single Schechter (dash-dotted lines in Fig. \ref{fig_U_LFs}), for comparison.

\subsubsection{Redshift evolution of the FUV, NUV and U-band LFs}

Figure \ref{fig_LF_evol} shows the redshift evolution of the fitted LF in UV and in the U-band at $0.05 < z \leq 3.5$ and $0.05 < z \leq 2.5$, respectively.
The first noteworthy feature is the evolution of the bright end of the LF, which fades continuously  with cosmic time (i.e., with decreasing redshift).
The second remarkable thing is the stability of the faint-end slopes in the FUV and NUV, which is clear up to $z \sim 1$. In other words, the populations of FUV and  NUV bright galaxies have been continuously decreasing since $z \sim 1$ while the populations of faint galaxies in these bands have remained stable. The same trends apply to the U-band LF evolution, although the faint end is noisier.   Additionally, it is interesting that the location of the upturn in the faint end slope of the U-band LF is preserved with cosmic time, in spite of the simultaneous recession of the bright end.

We can take this analysis further by considering how the values of the Schechter parameters change with redshift. 
Figure \ref{fig_LFparam_evol} shows the redshift evolution of $\alpha$ and $\phi^\star$ as a function of $M^\star$, corresponding to the fitted LFs shown in Fig. \ref{fig_LF_evol}.  The values of the slope $\alpha$ confirm the stability of the faint end seen in  Fig. \ref{fig_LF_evol} for the FUV and NUV LFs, with $-1.42 \leq \alpha \leq -1.31$ and $-1.53 \leq \alpha \leq -1.28$ at $0.05 < z \leq 1.3$, respectively.
The normalisation follows a similar trend with $\phi^\star = 4.4$--$6.0 \times 10^{-3}$ Mpc$^{-3}$ and $\phi^\star = 4.9$--$6.6 \times 10^{-3}$ Mpc$^{-3}$ at $0.05 < z \leq 1.3$ for the FUV and NUV LFs, respectively.  The characteristic absolute magnitude is characterized by a clear fading of $\sim$2.3 mag and $\sim$2 mag for $M^\star_\mathrm{FUV}$ and $M^\star_\mathrm{NUV}$, respectively, between $z \sim 2.2$ and  $z \sim 0.15$.

Regarding the U-band Schechter parameters, it is relevant to consider both the single and the double Schechter parametric forms, since our data are deep enough to allow us to observe a double-Schechter profile which has not been documented and discussed in the literature so far.  
When considering the single-Schechter parameters, the slope $\alpha$ and normalisation $\phi^\star$  appear particularly stable across cosmic time since $z \sim 1.6$, with $-1.4 \leq \alpha \leq -1.3$ and $\phi^\star = 3.0$--$5.0 \times 10^{-3}$ Mpc$^{-3}$; concurrently,  $M^\star_{\mathrm{U},\ \Sing}$ fades by $\sim$1.3 mag. However, a single Schechter may not be appropriate for the U-band LF, since -- as we discussed in the previous section -- the double Schechter appears to better fit the U-band LF measured at $0.05 < z \leq 1.8$.

When considering a double-Schechter, the dispersion observed in the faint end slope of the U-band LF is slightly larger, with $-1.70 \leq \alpha_1 \leq -1.44$ at $0.05 < z \leq 0.6$ (where we let $\alpha_1$ vary without any constraint except $\alpha_1 < \alpha_2$) and $-1.80 \leq \alpha_1 \leq -1.44$ at $0.05 < z \leq 1.8$. 
At the same time, the normalisation of the faint end appears stable, increasing slightly from $\phi^\star_1 = 1.0 \times 10^{-3}$ to $3.0 \times 10^{-3}$ Mpc$^{-3}$. 
At the bright end, $-0.60 \leq \alpha_2 \leq -0.20$ at $0.05 < z \leq 0.6$ and $-0.70 \leq \alpha_2 \leq -0.20$ at $0.05 < z \leq 1.8$, while $\phi^\star_2$ increases from $4.1 \times 10^{-3}$ to $7.3 \times 10^{-3}$ Mpc$^{-3}$ in the same redshift interval and the characteristic absolute magnitude $M^\star_{\mathrm{U},\ \Doub}$ fades by $\sim$1.7 mag.

While one may note that, on average, the double-Schechter parameters evolve in the same direction as the single-Schechter parameters, we note that the characteristic absolute magnitude depends substantially on the parametric form we adopted. In particular, the difference $|M^\star_{\mathrm{U},\ \Doub}-M^\star_{\mathrm{U},\ \Sing}|$ reaches $\sim0.8$ mag in our lowest redshift bin, i.e., where the faint-end upturn in the U-band LF is best probed.
Considering a double-Schechter fit of the U-band LF is therefore imperative to compare our estimation of $M^\star_\mathrm{U}$ with other estimates based on shallower surveys.

\begin{table*}
\begin{center}
\caption{Best-fit Schechter parameters of the FUV, NUV and U-band LFs and associated LDs.\label{table_param}}
\begin{tabular}{l*{9}{c}}
\multicolumn{9}{c}{FUV: single Schechter function} \\
\hline \\[-5mm]
\hline \\ [-3mm]
\multirow{2}{*}{Redshift} & \multicolumn{2}{c}{$^{*}$ $M_{lim}$ $^{(a)}$} & \multicolumn{2}{c}{$^{**}$ N$_{gal}^\mathrm{fit}$ ~~} & \multirow{2}{*}{$M^\star_\mathrm{FUV}$ $^{(a)}$} & \multirow{2}{*}{$\phi^\star$ $^{(b)}$} & \multirow{2}{*}{$\alpha$} & \multirow{2}{*}{$\log(\ \rho_\mathrm{FUV}$ $^{(c)} \ )$} \\
 & \begin{tiny}Deep\end{tiny} & \begin{tiny}Ultra-Deep\end{tiny} & \begin{tiny}Deep\end{tiny} & \begin{tiny}Ultra-Deep\end{tiny} & & & &  \\[0mm] 
\hline \\[-3mm]
$0.05 < z < 0.3$ & -14.21 & ----- & 111,819 & ----- &  -18.269$\pm0.054$ & 4.85$\pm0.35$ & -1.405$\pm0.019$ & 25.719$^{+0.009}_{-0.012}$ \\[1mm] 
$0.3 < z < 0.45$ & -15.17 & ----- & 150,738 & ----- &   -18.572$\pm0.038$ & 5.22$\pm0.25$ & -1.369$\pm0.017$ & 25.873$^{+0.005}_{-0.006}$ \\[1mm] 
$0.45 < z < 0.6$ & -15.79 & ----- & 168,370 & ----- &    -18.797$\pm0.073$ & 4.13$\pm0.50$ & -1.408$\pm0.053$ & 25.885$^{+0.012}_{-0.014}$ \\[1mm] 
$0.6 < z < 0.9$ & -17.16 & -16.34 & 265,051 &  17,888 &    -19.113$\pm0.041$ & 4.40$\pm0.30$ & -1.402$\pm0.038$ & 26.048$^{+0.009}_{-0.011}$ \\[1mm] 
$0.9 < z < 1.3$ & -18.02 & -17.27 & 422,362 &  28,890 &    -19.554$\pm0.065$ & 4.97$\pm0.57$ & -1.432$\pm0.068$ & 26.304$^{+0.018}_{-0.022}$ \\[1mm] 
$1.3 < z < 1.8$ & -18.79 & -17.80 & 305,245 &  36,289 &    -20.016$\pm0.074$ & 3.20$\pm0.38$ & -1.446$\pm0.074$ & 26.317$^{+0.022}_{-0.025}$ \\[1mm] 
$1.8 < z < 2.5$ & -19.67 & -18.91 & 180,033 &  22,979 &    -20.261$\pm0.042$ & 2.82$\pm0.11$ & -1.43 & 26.355$^{+0.005}_{-0.006}$ \\[1mm] 
$2.5 < z < 3.5$ & ----- & -19.73 & ----- &  54,089 &   -20.841$\pm0.046$ & 1.69$\pm0.10$ & -1.43 & 26.373$^{+0.011}_{-0.013}$ \\[1mm] 
\hline \\  
\end{tabular}   
\begin{tabular}{l*{9}{c}}
\multicolumn{9}{c}{NUV: single Schechter function} \\
\hline \\[-5mm]
\hline \\ [-3mm]
\multirow{2}{*}{Redshift} & \multicolumn{2}{c}{$^{*}$ $M_{lim}$ $^{(a)}$} & \multicolumn{2}{c}{$^{**}$ N$_{gal}^\mathrm{fit}$~~} & \multirow{2}{*}{$M^\star_\mathrm{NUV}$ $^{(a)}$} & \multirow{2}{*}{$\phi^\star$ $^{(b)}$} & \multirow{2}{*}{$\alpha$} & \multirow{2}{*}{$\log(\ \rho_\mathrm{NUV}$ $^{(c)} \ )$} \\
 & \begin{tiny}Deep\end{tiny} & \begin{tiny}Ultra-Deep\end{tiny} & \begin{tiny}Deep\end{tiny} & \begin{tiny}Ultra-Deep\end{tiny} & & & &  \\[0mm] 
\hline \\[-3mm]
$0.05 < z < 0.3$ & -14.38 & -----  & 117,326 & ----- &   -18.514$\pm0.025$ & 5.08$\pm0.2$ & -1.399$\pm0.011$ & 25.847$^{+0.006}_{-0.006}$ \\[1mm] 
$0.3 < z < 0.45$ & -15.37 & -----  & 157,999 & ----- &   -18.798$\pm0.03$ & 6.01$\pm0.26$ & -1.308$\pm0.014$ & 26.009$^{+0.007}_{-0.008}$ \\[1mm] 
$0.45 < z < 0.6$ & -16.02 & -----  & 174,053 & ----- &   -19.026$\pm0.062$ & 4.56$\pm0.47$ & -1.364$\pm0.043$ & 26.009$^{+0.009}_{-0.013}$ \\[1mm] 
$0.6 < z < 0.9$ & -17.27 & -16.49 & 297,674 &  17,545 &   -19.416$\pm0.053$ & 4.24$\pm0.4$ & -1.396$\pm0.044$ & 26.159$^{+0.008}_{-0.010}$ \\[1mm] 
$0.9 < z < 1.3$ & -18.25 & -17.49 & 404,623 &  28,462 &   -19.859$\pm0.052$ & 4.7$\pm0.43$ & -1.385$\pm0.046$ & 26.385$^{+0.009}_{-0.01}$ \\[1mm] 
$1.3 < z < 1.8$ & -19.13 & -18.14 & 293,618 &  34,610 &   -20.367$\pm0.045$ & 3.13$\pm0.23$ & -1.391$\pm0.038$ & 26.422$^{+0.006}_{-0.008}$ \\[1mm] 
$1.8 < z < 2.5$ & -20.05 & -19.24  & 173,206 &  23,007 &   -20.622$\pm0.034$ & 2.72$\pm0.08$ & -1.4 & 26.472$^{+0.004}_{-0.004}$ \\[1mm] 
$2.5 < z < 3.5$ & ----- & -20.15 & ----- & 23,598 &   -21.152$\pm0.038$ & 1.71$\pm0.11$ & -1.4 & 26.489$^{+0.015}_{-0.017}$ \\[1mm]  
\hline \\
\end{tabular}
\begin{tabular}{l*{9}{c}}
\multicolumn{9}{c}{U-band: single Schechter function} \\
\hline \\[-5mm]
\hline \\ [-3mm]
\multirow{2}{*}{Redshift} & \multicolumn{2}{c}{$^{*}$ $M_{lim}$ $^{(a)}$} & \multicolumn{2}{c}{$^{**}$ N$_{gal}^\mathrm{fit}$~~} & \multirow{2}{*}{$M^\star_{\mathrm{U},\ \Sing}$ $^{(a)}$} & \multirow{2}{*}{$\phi^\star$ $^{(b)}$} & \multirow{2}{*}{$\alpha$} & \multirow{2}{*}{$\log(\ \rho_{\mathrm{U},\ \Sing}$ $^{(c)} \ )$} \\
 & \begin{tiny}Deep\end{tiny} & \begin{tiny}Ultra-Deep\end{tiny} & \begin{tiny}Deep\end{tiny} & \begin{tiny}Ultra-Deep\end{tiny} & & & &  \\[0mm] 
\hline \\[-3mm]
$0.05 < z < 0.3$ & -15.17 & ----- & 121,413 & ----- &   -19.865$\pm0.030$ & 3.60$\pm0.12$ & -1.424$\pm0.007$ & 26.291$^{+0.004}_{-0.005}$ \\[1mm] 
$0.3 < z < 0.45$ & -16.20 & ----- & 155,008 & ----- &   -20.042$\pm0.020$ & 5.62$\pm0.13$ & -1.22$\pm0.008$ & 26.466$^{+0.003}_{-0.003}$ \\[1mm] 
$0.45 < z < 0.6$ & -16.86 & ----- & 166,133 & ----- &   -20.125$\pm0.017$ & 5.05$\pm0.11$ & -1.178$\pm0.009$ & 26.439$^{+0.002}_{-0.002}$ \\[1mm] 
$0.6 < z < 0.9$ & -17.91 &  -17.22 & 356,389 &  13,958 &   -20.435$\pm0.012$ & 5.28$\pm0.08$ & -1.154$\pm0.008$ & 26.576$^{+0.002}_{-0.002}$ \\[1mm] 
$0.9 < z < 1.3$ & -18.92 &  -18.21 & 461,661 &  24,246 &   -20.841$\pm0.014$ & 4.66$\pm0.09$ & -1.251$\pm0.01$ & 26.723$^{+0.002}_{-0.002}$ \\[1mm] 
$1.3 < z < 1.8$ & -19.74 &  -18.85 & 354,083 &  29,295 &   -21.241$\pm0.017$ & 2.96$\pm0.07$ & -1.352$\pm0.013$ & 26.737$^{+0.004}_{-0.004}$ \\[1mm] 
$1.8 < z < 2.5$ &  ----- &  -20.32 & ----- &  30,689 &   -21.677$\pm0.039$ & 2.2$\pm0.12$ & -1.394$\pm0.033$ & 26.808$^{+0.011}_{-0.011}$ \\[1mm] 
\hline \\[-2mm]
\end{tabular}
\begin{tabular}{l*{7}{c}}
\multicolumn{7}{c}{U-band: double Schechter function} \\
\hline \\[-5mm]
\hline \\ [-3mm]
Redshift  & $M^\star_{\mathrm{U},\ \Doub}$ $^{(a)}$ & $\phi^\star_1$ $^{(b)}$ & $\alpha_1$ & $\phi^\star_2$ $^{(b)}$ & $\alpha_2$ & $\log(\ \rho_{\mathrm{U},\ \Doub}$ $^{(c)} \ )$\\[0mm] 
\hline \\[-3mm]
$0.05 < z < 0.3$ &   -18.961$\pm0.050$ & 3.4$\pm0.36$ & -1.568$\pm0.024$ & 7.16$\pm0.25$ & -0.213$\pm0.099$ & 26.301$^{+0.003}_{-0.003}$ \\[1mm] 
$0.3 < z < 0.45$ &   -19.500$\pm0.040$ & 2.41$\pm0.47$ & -1.557$\pm0.051$ & 7.91$\pm0.3$ & -0.419$\pm0.092$ & 26.479$^{+0.002}_{-0.003}$ \\[1mm] 
$0.45 < z < 0.6$ &   -19.744$\pm0.027$ & 1.96$\pm0.08$ & -1.56 & 5.59$\pm0.12$ & -0.506$\pm0.043$ & 26.452$^{+0.002}_{-0.002}$ \\[1mm] 
$0.6 < z < 0.9$ &   -20.244$\pm0.023$ & 1.46$\pm0.11$ & -1.56& 5.04$\pm0.09$ & -0.773$\pm0.042$ & 26.588$^{+0.002}_{-0.002}$ \\[1mm] 
$0.9 < z < 1.3$ &   -20.819$\pm0.033$ & 0.51$\pm0.63$ & -1.56 & 4.28$\pm0.47$ & -1.187$\pm0.092$ & 26.727$^{+0.014}_{-0.001}$ \\[1mm] 
\hline \\[0mm] 
\multicolumn{7}{l}{\begin{footnotesize} $^{*}~$ Absolute magnitude completeness limits of the Deep and Ultra-Deep layers (see Equation~\ref{eq_Mabs_lim_eff}).  \end{footnotesize}} \\
\multicolumn{7}{l}{\begin{footnotesize} $^{**}$ Number of galaxies with $M < M_{lim}$ from the Deep and Ultra-Deep layers used for fitting.  \end{footnotesize}} \\
\multicolumn{7}{l}{\begin{footnotesize} $^{(a)}$  AB mag.  \end{footnotesize}} \\
\multicolumn{7}{l}{\begin{footnotesize} $^{(b)}$  $10^{-3}$ Mpc$^{-3}$.  \end{footnotesize}} \\
\multicolumn{7}{l}{\begin{footnotesize} $^{(c)}$  erg  Hz$^{-1}$ s$^{-1}$ Mpc$^{-3}$.  \end{footnotesize}} \\
\end{tabular}
\end{center}
\end{table*}

\subsubsection{Comparison with previous studies}

\begin{figure*}
\center
\includegraphics[width=\hsize]{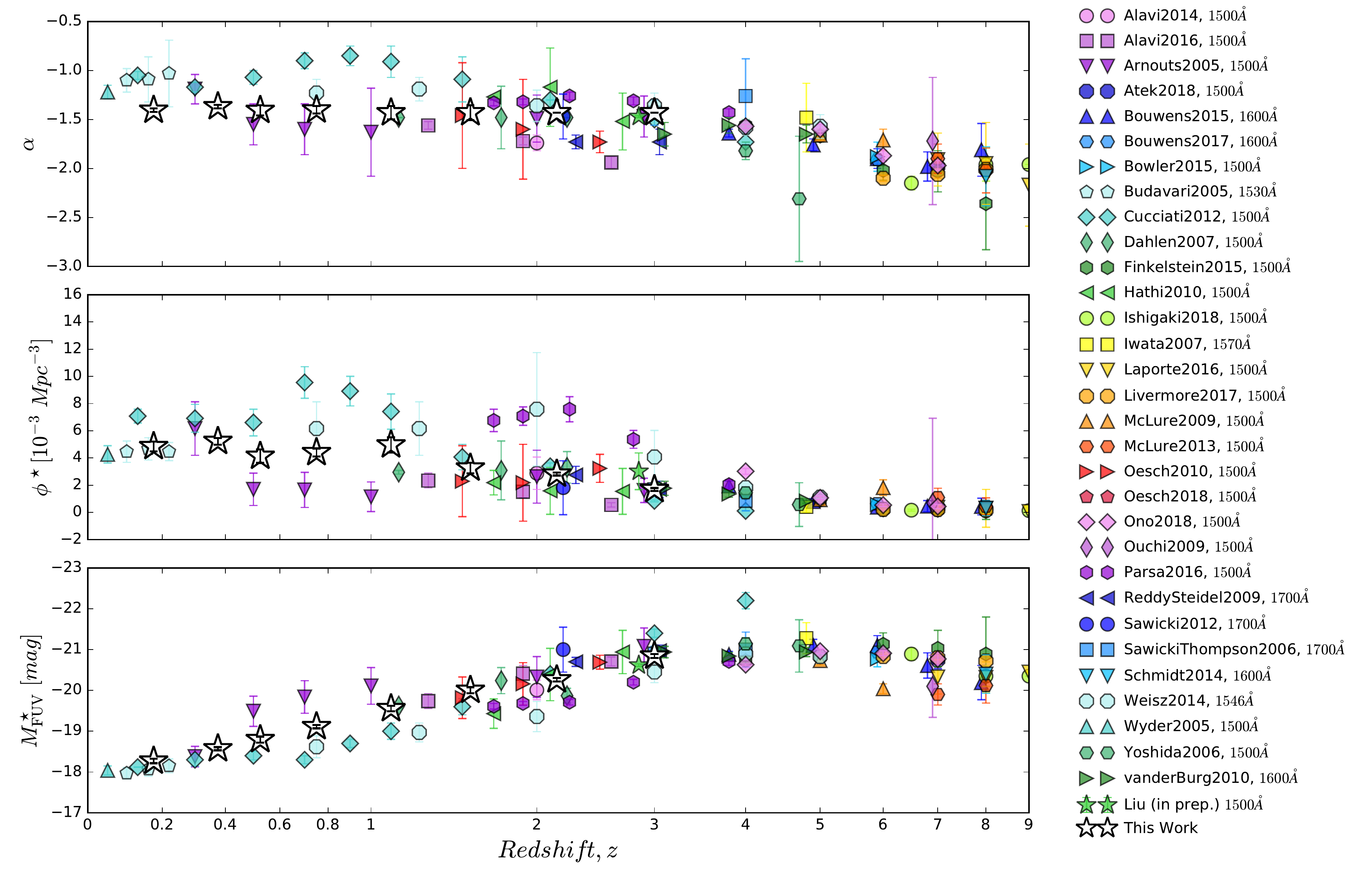}
\caption{Comparison of the FUV Schechter parameters we found (open stars) with values compiled from the literature, namely, 
\citet{Alavi2014}, 
\citet{Alavi2016},
\citet{Arnouts2005},
\citet{Atek2018},
\citet{Bouwens2015a},
\citet{Bouwens2017},
\citet{Bowler2015},
\citet{Budavari2005},
\citet{Cucciati2012},
\citet{Dahlen2007},
\citet{Finkelstein2015},
\citet{Hathi2010},
\citet{Ishigaki2018},
\citet{Iwata2007},
\citet{Laporte2016},
\citet{Livermore2017},
\citet{McLure2009},
\citet{McLure2013},
\citet{Oesch2010},
\citet{Oesch2018},
\citet{Ono2018},
\citet{Ouchi2009},
\citet{Parsa2016},
\citet{ReddySteidel2009},
\citet{SawickiThompson2006a},
\citet{Sawicki2012},
\citet{Schmidt2014},
\citet{Weisz2014},
\citet{Wyder2005},
\citet{Yoshida2006},
\citet{vanderBurg2010}, and C. Liu et al. (in prep.; light green stars) who used the same CLAUDS+HSC observations we used but based their LF measurements on u-dropouts.
\label{fig_FUV_LFparam_lit}  }
\end{figure*}

Often simply referred to as the UV LF, the FUV LF has been extensively studied up to redshift $z \sim 9$, where rest-frame (not dust-corrected) UV can be constrained from mid-infrared observations. In Fig. \ref{fig_FUV_LFparam_lit}, we compare our best-fit Schechter parameters for the FUV LF with values from the literature across redshift. As one can see, our results are in overall good agreement with the literature at $0.05 < z \leq 3.5$.  Given its unsurpassed combination of depth and area, our homogeneous dataset provides the definitive reference measurement of the rest-frame FUV LF out to $z\sim3$ at this time. 

It is remarkable how well-behaved the values of the Schechter parameters are with redshift in Fig.~\ref{fig_FUV_LFparam_lit} over the redshift range we measured them:  $M^\star$ increases monotonically with lookback time, while both  $\alpha$ and $\phi^\star$ remain essentially constant.  The FUV LF slope $\alpha$, which we measured directly from $z \sim 1.6$, is of particular interest as it appears flatter than what is reported in the literature higher redshifts, from $z \sim 9$. This points to the existence of two regimes in the evolution of the FUV LF's faint end, which flattened from $z \sim 9$ before stabilizing at or before  $z \sim 1.6$. Similarly, the evolution of $\phi^\star$ we measure is very stable from $z \sim 1.6$, and seem to be right in the middle of the literature values. 
At the same time, the continuous fading of the FUV LF's bright end characteristic absolute magnitude, $M^\star$  we observe from $z \sim 3$ is in line with the literature, though much better constrained with our data.

\begin{figure*}
\center
\includegraphics[width=0.46\hsize]{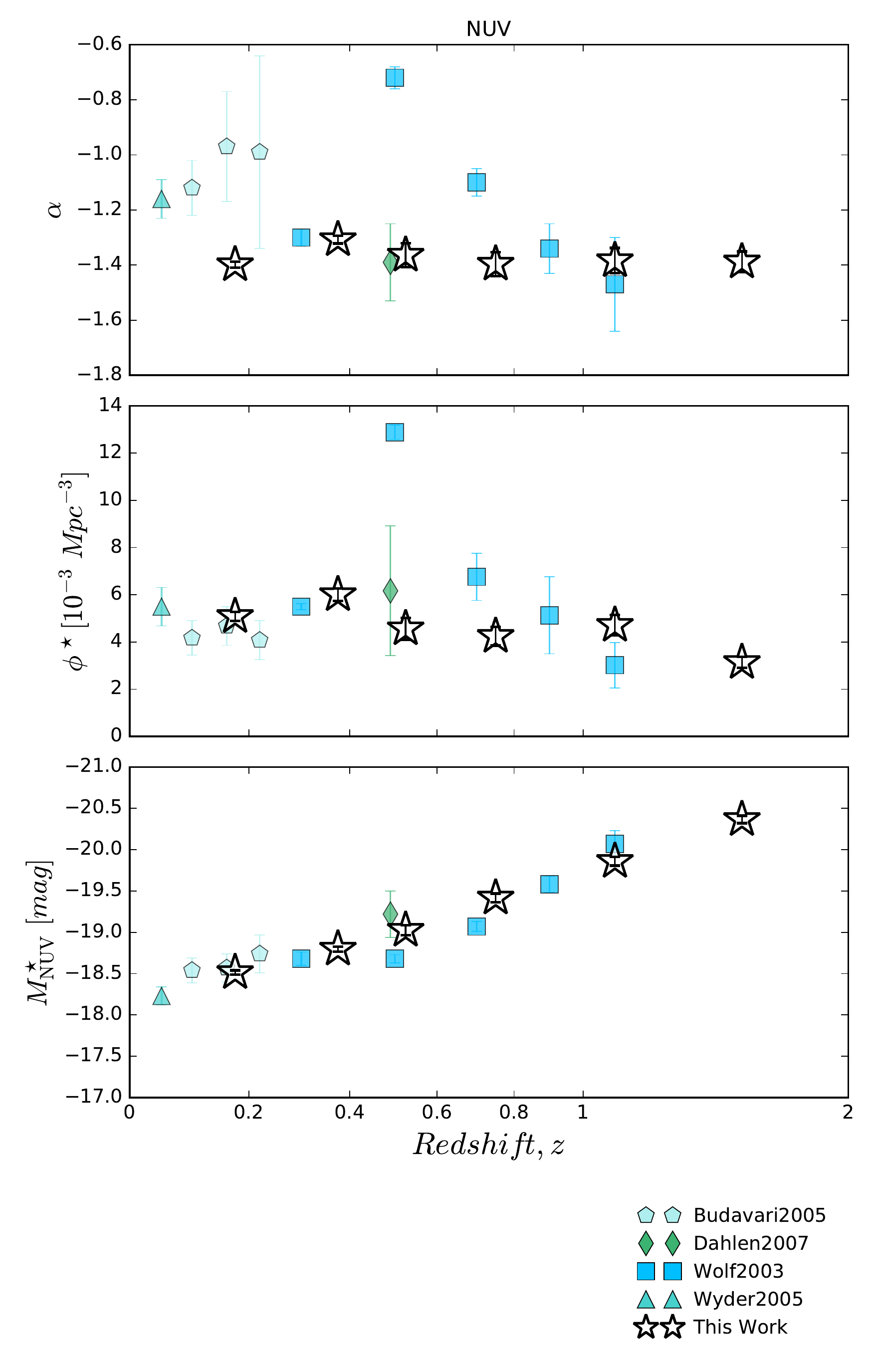}
\includegraphics[width=0.46\hsize]{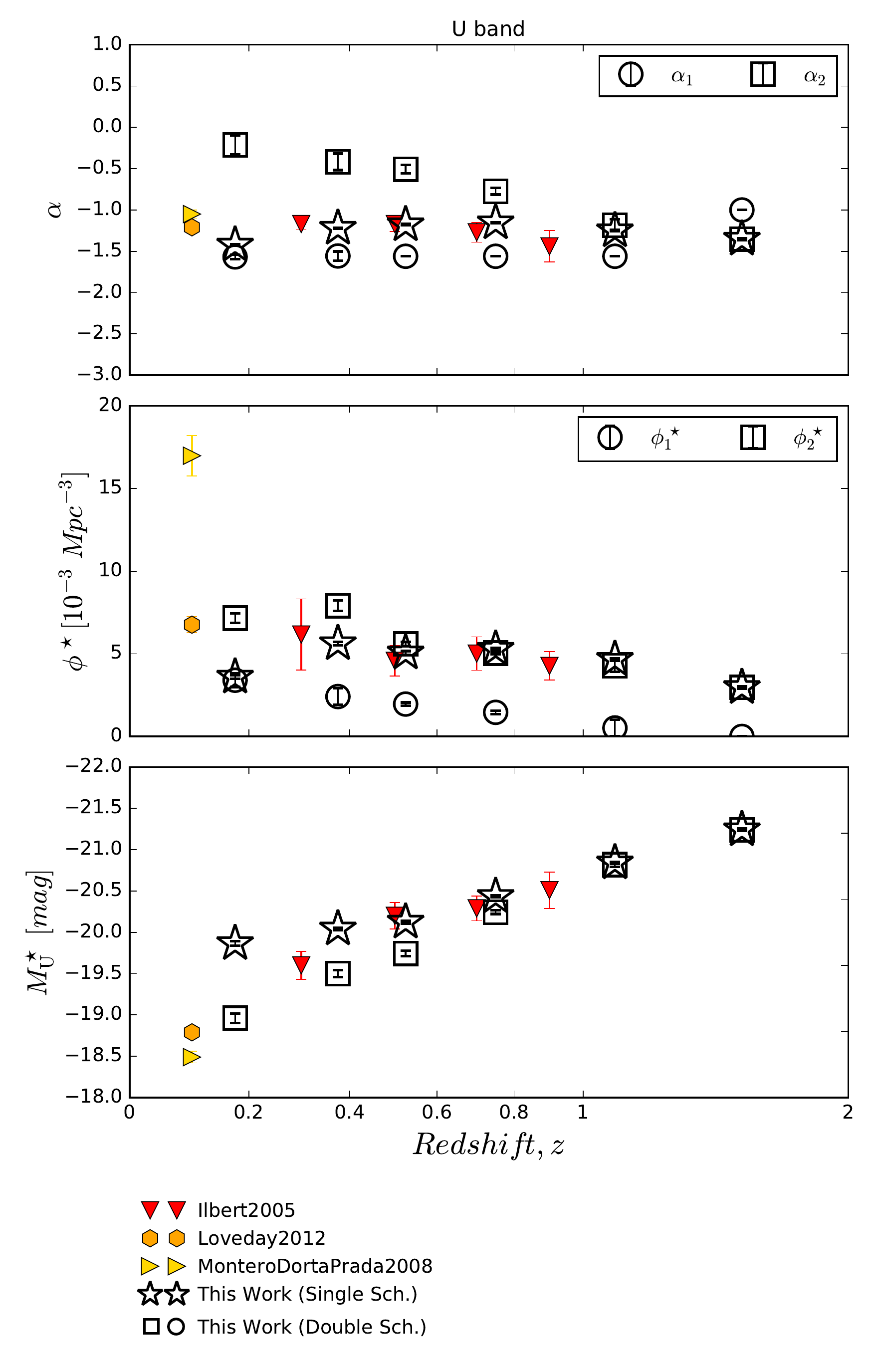}
\caption{Comparison of the NUV and U-band (single-)Schechter parameters (open stars) with the literature, namely \citet{Budavari2005},
\citet{Dahlen2007},
\citet{Wolf2003},
\citet{Wyder2005},
\citet{Ilbert2005},
\citet{Loveday2012}, and
\citet{MonteroDortaPrada2009}. 
In the right-hand panel, the double-Schechter best-fitting parameters we obtained for the U-band LF are over-plotted (open circles and squares).
\label{fig_NUV_U_LFparam_lit}  }
\end{figure*}

In contrast to the FUV LF, the NUV and U-band LFs are much less well documented in the literarture, especially at $ z > 1.5$. Figure \ref{fig_NUV_U_LFparam_lit} shows our measurements of the NUV and U-band Schechter parameters  compared with those from the literature. For the literature compilation we only considered analyses where the parameters were free to vary over the redshift range covered by the literature, i.e., up to $z \sim 1.5$.  

Our NUV LF Schechter parameters are in overall good agreement with the literature, but provide measurements that are much less noisy. This is particularly clear for the redshift dependence of $\alpha$ (and to a lower extent for $\phi^\star$ ), for which our measurement is  more stable than what is found the literature. Our $\alpha$ values, in particular, show a remarkable stability with redshift.  At the same time, the evolution of $M^\star_\mathrm{NUV}$ we measured is in very good agreement with the literature, although even less noisy; with our excellent statistics, it shows a remarkably steady progression with cosmic time. 

For the U-band LF, the comparison with the literature is different if we consider the single or double Schechter function fit. When considering a single-Schechter (star sybmols in Fig. \ref{fig_NUV_U_LFparam_lit}), the agreement with the literature is particularly good, especially for $\alpha$ and  $\phi^\star$, while one may notice a little discrepancy for $M^\star_\mathrm{U}$ at $z < 0.5$. This is expected as the faint-end excess of galaxies in the U-band LF is more pronounced at low redshift, which directly affects our estimation of $M^\star_\mathrm{U}$ due to the well known degeneracy between the Schechter parameters $\alpha$ and $M^\star$ (as observed in Fig. \ref{fig_LFparam_evol}).
Thus, $M^\star_{\mathrm{U}, \Doub}$ is in overall good agreement with the literature from $z \sim 0$ up to $z \sim 1$ (i.e., over all the redshift range where the comparison is possible), while exhibiting a much less noisy evolution.

\subsection{Redshift evolution of the luminosity densities}

\subsubsection{Luminosity density from the LF}
\label{sect_LDmeasurement}

In principle, the luminosity density (LD) is obtained by summing the light from all the galaxies in unit volume. In practice, the LD can be estimated by integrating the LF. The luminosity density of galaxies with luminosity greater than $L$ is defined by 
\begin{equation}
\rho(L) = \int_{L}^{\infty} L'\ \phi(L') \ dL' ~. 
\label{eq_LDintegral}
\end{equation}
If the LF has the (single) Schechter form, this reduces to
\begin{equation}
\rho(L) = \phi^\star \ L^\star \ \Gamma(\alpha+2, L/L^\star)
\label{eq_LD_oneSchechter}
\end{equation}
where $\Gamma$ is upper incomplete gamma function. In the case of a double-Schechter LF, Equation~\ref{eq_LDintegral} becomes
\begin{equation}
\rho(L) = L^\star \ \left[ \ \phi_1^\star  \ \Gamma(\alpha_1+2, L/L^\star) + \ \phi_2^\star  \ \Gamma(\alpha_2+2, L/L^\star) \ \right] ~.
\label{eq_LD_twoSchechter}
\end{equation}

We derive the rest-frame FUV, NUV, and U-band LDs using the Schechter parameters we obtained in Sec.~\ref{sect_LFmeasurements} and Equations~\ref{eq_LD_oneSchechter} or \ref{eq_LD_twoSchechter}, as 
appropriate. We integrate over luminosity from $\infty$ down to 
$M_\mathrm{FUV},M_\mathrm{NUV},M_\mathrm{U} = -15$ to avoid heavy extrapolations. This limit
is $\geq3$ magnitudes below $M^\star$ for all of our LF measurements, 
and -- given our relatively shallow values of $\alpha$ -- it therefore captures the vast bulk of the luminosity that escapes the galaxy population. 
We present the resulting LD values in the last column of Table \ref{table_param} and discuss the results in the next section. 


\subsubsection{Redshift evolution of the FUV, NUV and U-band LDs}

\begin{figure*}
\center
\includegraphics[width=\hsize, trim =0.5cm 0.4cm 0.7cm 0cm, clip]{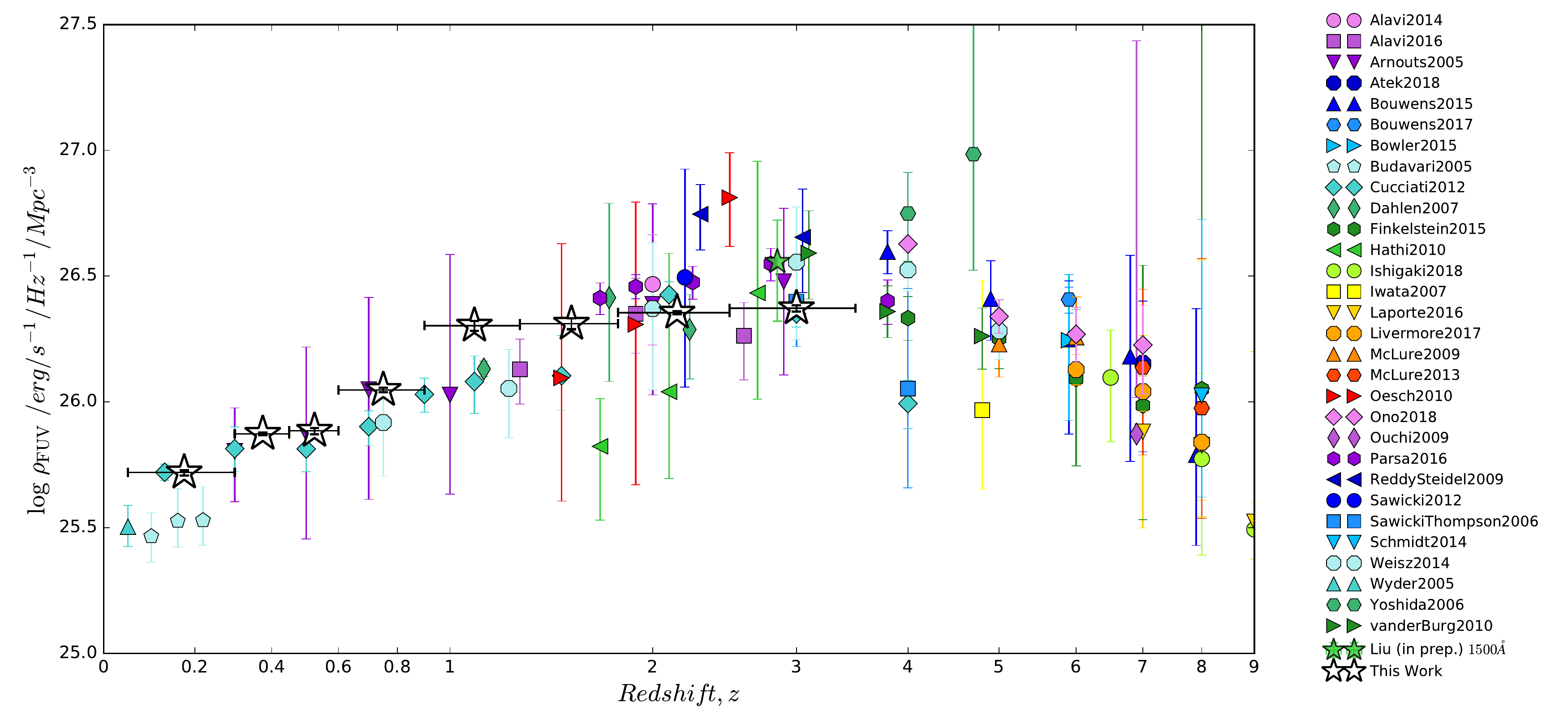}
\caption{Redshift evolution of the FUV luminosity densities for galaxies brighter than $M_\mathrm{FUV} = -15$ (open stars), and comparison with values derived from the literature, namely, 
\citet{Alavi2014}, 
\citet{Alavi2016},
\citet{Arnouts2005},
\citet{Atek2018},
\citet{Bouwens2015a},
\citet{Bouwens2017},
\citet{Bowler2015},
\citet{Budavari2005},
\citet{Cucciati2012},
\citet{Dahlen2007},
\citet{Finkelstein2015},
\citet{Hathi2010},
\citet{Ishigaki2018},
\citet{Iwata2007},
\citet{Laporte2016},
\citet{Livermore2017},
\citet{McLure2009},
\citet{McLure2013},
\citet{Oesch2010},
\citet{Oesch2018},
\citet{Ono2018},
\citet{Ouchi2009},
\citet{Parsa2016},
\citet{ReddySteidel2009},
\citet{SawickiThompson2006a},
\citet{Sawicki2012},
\citet{Schmidt2014},
\citet{Weisz2014},
\citet{Wyder2005},
\citet{Yoshida2006},
\citet{vanderBurg2010}, and C. Liu et al. (in prep.; light green stars).
\label{fig_FUV_LD_evol}  }
\end{figure*}

\begin{figure*}
\center
\includegraphics[width=0.46\hsize, trim =0.35cm 1cm 0cm 0.5cm, clip]{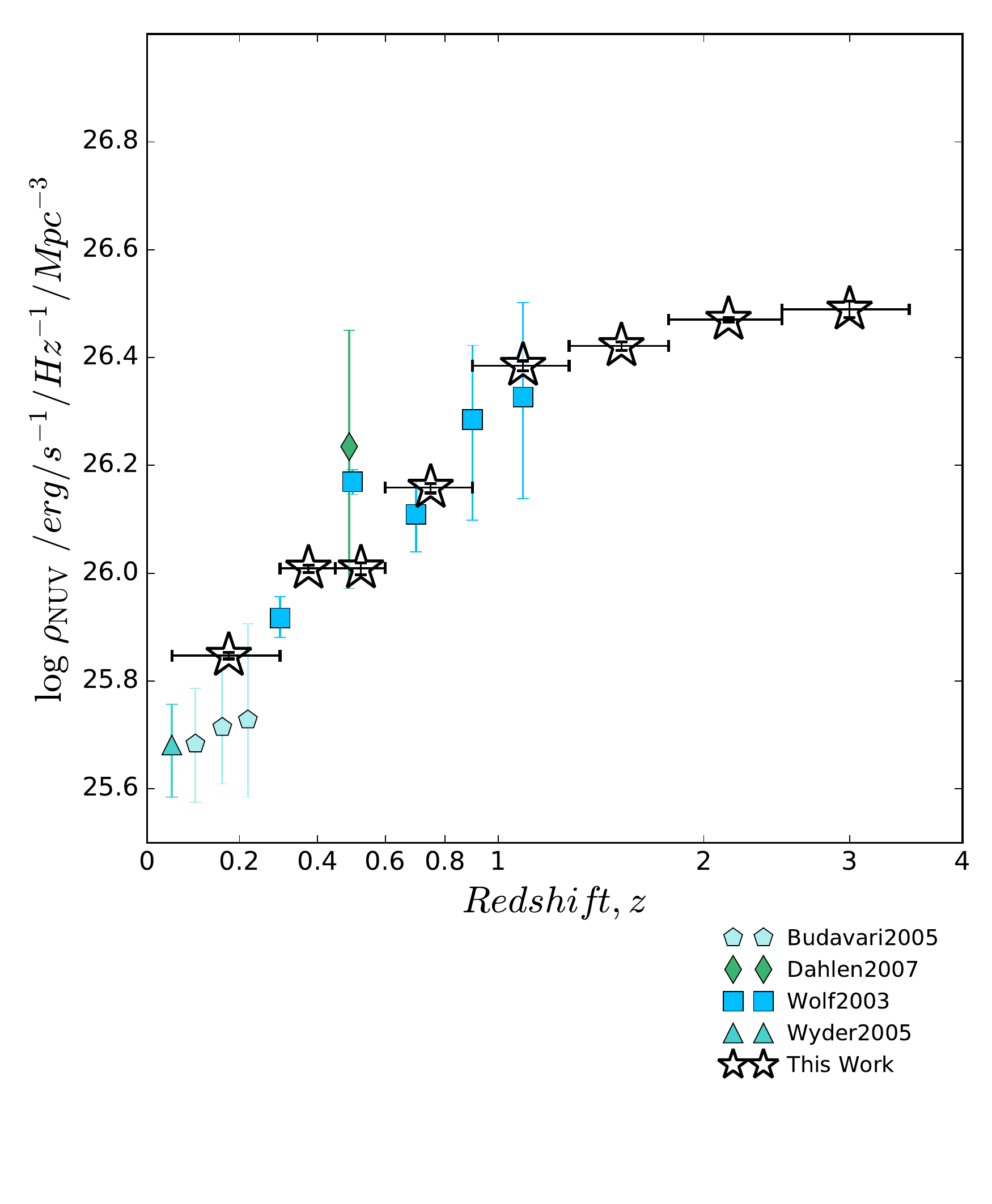}
\includegraphics[width=0.46\hsize, trim =0cm 1cm 0.35cm 0.5cm, clip]{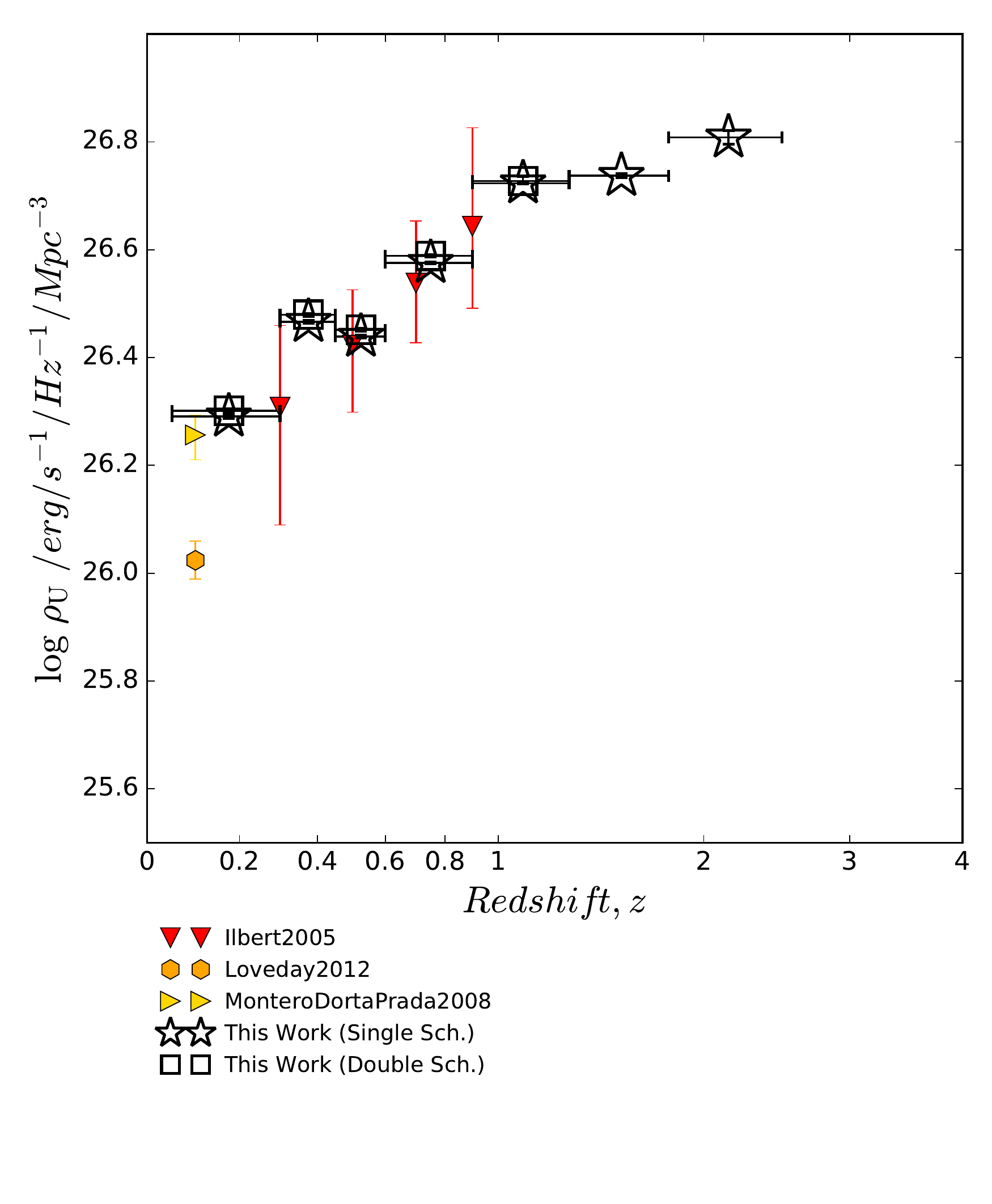}
\caption{Redshift evolution of the NUV and U-band luminosity densities for galaxies brighter than $M_\mathrm{NUV},M_\mathrm{U} = -15$ (open stars), and comparison with values derived from the literature, namely \citet{Budavari2005},
\citet{Dahlen2007},
\citet{Wolf2003},
\citet{Wyder2005},
\citet{Ilbert2005},
\citet{Loveday2012}, and
\citet{MonteroDortaPrada2009}.
In the right-hand panel, the U-band LD we derived assuming a single ($\rho_{\mathrm{U},\ \Sing}$; open stars) and double ($\rho_{\mathrm{U},\ \Doub}$; open squares) Schechter parametric forms are reported.
\label{fig_NUV_U_LD_evol}  }
\end{figure*}

Figures \ref{fig_FUV_LD_evol} and \ref{fig_NUV_U_LD_evol} show the redshift evolution of our FUV, NUV and U-band luminosity densities measured as described in
Sec.~\ref{sect_LDmeasurement}.  For comparison, we show LD values we recalculated from literature LF measurements for the same luminosity limits as those we applied to the CLAUDS+HSC-SSP data. .

In Fig. \ref{fig_FUV_LD_evol} one can see how our results support a picture where the FUV luminosity density has continuously decreased from $\rho_\mathrm{FUV} \sim 10^{26.35}$ down to $\sim 10^{25.7}$ erg s$^{-1}$ Hz$^{-1}$ Mpc$^{-3}$ between $z \sim 2$ and $z \sim 0.2$, in good agreement with the literature. At the same time, our results show $\rho_\mathrm{FUV}$ to be stable at $1 \lesssim z \lesssim 2$ (and even at $1 \lesssim z \lesssim 3$ if we assume that the slope of $\alpha=-1.43$ we have set at $z > 1.8$ from lower-$z$ measurements is correct).  In that respect, our results appear to be consistent with a picture where the cosmic UV luminosity density experienced a relatively stable phase before decreasing exponentially from redshift $z \sim 1$.

In Fig. \ref{fig_NUV_U_LD_evol}a, one can see a similar trend for the redshift evolution of the NUV luminosity density, with a continuous decrease from $\rho_\mathrm{NUV} \sim 10^{26.4}$ down to $\sim 10^{25.85}$ erg s$^{-1}$ Hz$^{-1}$ Mpc$^{-3}$ between $z \sim 1$ and $z \sim 0.2$, after a less pronounced evolution at $1 \lesssim z \lesssim 2$ (and also at $1 \lesssim z \lesssim 3$, assuming a slope of $\alpha=-1.4$ at $z > 1.8$). 
The  evolution of the U-band luminosity density shows a similar trend, at least at $z < 2$, with a continuous decrease of from $\rho_\mathrm{U} \sim 10^{26.8}$ down to $\sim 10^{26.25}$ erg s$^{-1}$ Hz$^{-1}$ Mpc$^{-3}$ between $z \sim 2$ and $z \sim 0.2$ and a more stable evolution at $1 \lesssim z \lesssim 2$, irrespective of whether we consider the double- or single-Schechter fits of the U-band LF, with the difference $\vert \log( \rho_{\mathrm{U},\ \Doub}) - \log( \rho_{\mathrm{U},\ \Sing}) \ \vert \lesssim~0.01$~dex. 
That is in fairly good agreement with the literature shown in Figs. \ref{fig_NUV_U_LD_evol}a and b, although comparison is only possible up to $z \sim 1$ for $\rho_\mathrm{NUV}$ and $\rho_\mathrm{U}$.  These results are in broad agreement with the picture first presented by \cite{Sawicki1997}, namely that of a broad plateau at $1 \lesssim z \lesssim 3.5$ followed by a steep decline from $z\sim1$ to $z\sim 0$ \citep[the latter first measured by][]{Lilly1996}.\footnote{The \cite{Sawicki1997} measurements were performed at rest-frame UV wavelengths but then were extrapolated to rest-frame 3000\AA\ (roughly mid-way between the NUV and U-band of the present study) for homogeneity with the $z \lesssim 1$ measurements of \cite{Lilly1996}.} 

Our CLAUDS+HSC-SSP measurements (Sec.~\ref{sect_results}) show that the evolution of the FUV and NUV LDs out to $z\sim 1$ is primarily driven by changes in $M^\star$ rather than in the faint-end slope, $\alpha$, or the number density of galaxies, $\phi^\star$. At higher redshifts, $z \gtrsim 1$, while $M^\star$ continues to brighten, $\phi^\star$ begins to drop, with the two effects balancing each other to give the much milder, evolution seen at $z \gtrsim 1$ in Figs.~\ref{fig_FUV_LD_evol} and \ref{fig_NUV_U_LD_evol}a.  The interpretation is more complicated in the rest-frame U-band because of the double-Schechter form of the U-band LF.  There, we suspect that the build-up of the population of quiescent galaxies may contribute to the LF (bright end) and LD, as we explore in a forthcoming companion paper (T.~Moutard et al., in prep.).

\section{Summary}

In this paper we presented our measurements of the $0 < z \lesssim 3$ rest-frame FUV (1546\AA), NUV (2345\AA), and U-band (3690\AA) galaxy luminosity functions and luminosity densities using 
more than 4.3 million galaxies from the CLAUDS and HSC-SSP surveys. 
The unprecedented combination of depth (U$\sim$27) and area ($\sim$18deg$^2$) of this dataset allows us to constrain the shape and evolution of these LFs with unmatched statistical precision and essentially free of cosmic variance. 

The main results of this paper are the LF and LD measurements presented in the Figures and Tables in Section~\ref{sect_results}.  In addition to these main products, we wish to highlight again the following observations: 

\begin{enumerate}
    \item The rest-frame FUV and NUV luminosity functions are described very well by the classic Schechter form over the full redshift range studied.  The evolution of the Schechter parameters is very smooth with redshift:  In particular, the values of $M^\star$ for both the FUV and NUV increase monotonically with increasing redshift, while the faint-end slopes are very stable up to $z \sim 2$, with slope values conservatively within $-1.42 \leq \alpha_{FUV} \leq -1.31$ and $-1.53 \leq \alpha_{NUV} \leq -1.28$ over $0.05 < z \leq 1.3$. 
    
    \item In contrast to the FUV and NUV LFs, the rest-frame U-band luminosity functions are best described by a double Schechter model, $M^\star_{\mathrm{U},\ \Doub}$, $\phi^\star_\mathrm{U,1}$,  $\phi^\star_\mathrm{U,2}$, and $\alpha_\mathrm{U,1}$ evolving continuously through $0.2<z<2$, assuming that $\alpha_\mathrm{U,2}$ is simultaneously stable with redshift, which is confirmed to at least $z\sim0.5$ (we are unable to measure it independently beyond this redshift).  We speculate that the second Schechter component in the rest-frame U-band LF is due to the population of quiescent galaxies -- a topic we are currently investigating in a companion paper (T.~Moutard, in prep.). 
    
    \item We measured the rest-frame FUV, NUV, and U-band luminosity densities by integrating the corresponding LFs down to $M=-15$ at $z \sim 0.2$. At all three wavelengths we confirm previous results but with much better statistical precision afforded by our wide-and-deep CLAUDS+HSC-SSP dataset:  at all three rest wavelengths the luminosity density increases monotonically and rapidly with lookback time from $z \sim 0.2$ to $z \sim 1$ and then flattens to a much gentler slope at $z > 1$.  
    
    \item The very shallow evolution of the FUV and NUV LDs from $ z \sim 3$ to $z \sim 1$ is driven by two competing effects acting within the LFs:  the fading of the characteristic magnitude $M^\star$, which is balanced by the increase in the number of objects, $\phi^\star$ to produce the essentially flat LDs we observe over this wavelength range.  At $z<1$ the rapid evolution of the luminosity densities is essentially due to the continuing fading of $M^\star$ only as both $\phi^\star$ and $\alpha$ remain essentially constant from $z \sim 1$ to $z \sim 0.2$.
    
\end{enumerate}

We hope that the $0<z<3$ LF and LD  measurements we presented in this paper  will serve as a useful reference to the community  for making observational forecasts and validating theoretical models.   In the future, we plan to extend the range of our LF and LD measurements to higher redshifts (0<z<7) by incorporating Lyman Break Galaxy luminisity functions that we plan to do in a consistent way across this redshift range.

\section*{Acknowledgements}

We gratefully acknowledge the anonymous reviewer, whose insightful comments helped in improving the clarity of the paper.
We thank the CFHT observatory staff for their hard work in obtaining these data. The observations presented here were performed with care and respect from the summit of Maunakea which is a significant cultural and historic site.  We thank Guillaume Desprez and Chengze Liu for helpful suggestions. 

This work is based on observations obtained with MegaPrime/ MegaCam, a joint project of CFHT and CEA/DAPNIA, at the Canada-France-Hawaii Telescope (CFHT) which is operated by the National Research Council (NRC) of Canada, the Institut National des Science de l'Univers of the Centre National de la Recherche Scientifique (CNRS) of France, and the University of Hawaii. This research uses data obtained through the Telescope Access Program (TAP), which has been funded by the National Astronomical Observatories, Chinese Academy of Sciences, and the Special Fund for Astronomy from the Ministry of Finance. This work uses data products from TERAPIX and the Canadian Astronomy Data Centre. It was carried out using resources from Compute Canada and Canadian Advanced Network For Astrophysical Research (CANFAR) infrastructure. These data were obtained and processed as part of CLAUDS, which is a collaboration between astronomers from Canada, France, and China described in \cite{Sawicki2019}.

This work is also based in part on data collected at the Subaru Telescope and retrieved from the HSC data archive system, which is operated by the Subaru Telescope and Astronomy Data Center at National Astronomical Observatory of Japan. The Hyper Suprime-Cam (HSC) collaboration includes the astronomical communities of Japan and Taiwan, and Princeton University.  The HSC instrumentation and software were developed by the National Astronomical Observatory of Japan (NAOJ), the Kavli Institute for the Physics and Mathematics of the Universe (Kavli IPMU), the University of Tokyo, the High Energy Accelerator Research Organization (KEK), the Academia Sinica Institute for Astronomy and Astrophysics in Taiwan (ASIAA), and Princeton University.  Funding was contributed by the FIRST program from Japanese Cabinet Office, the Ministry of Education, Culture, Sports, Science and Technology (MEXT), the Japan Society for the Promotion of Science (JSPS),  Japan Science and Technology Agency  (JST),  the Toray Science  Foundation, NAOJ, Kavli IPMU, KEK, ASIAA,  and Princeton University. This paper makes use of software developed for the Large Synoptic Survey Telescope. We thank the LSST Project for making their code available as free software at http://dm.lsst.org. 

This work was financially supported by a Discovery Grant from the Natural Sciences and Engineering Research Council (NSERC) of Canada, by the Programme National Cosmology et Galaxies (PNCG) of CNRS/INSU with INP and IN2P3, and by the Centre National d'Etudes Spatiales (CNES).




\bibliographystyle{mnras}
\bibliography{Moutard2019_CLAUDS_LF.bib}





\newpage

\appendix

\section{Fitting the U-band luminosity function}
\label{app_U_LF_fitting}

\begin{figure*}
\center
\includegraphics[width=\hsize, trim = 0.5cm 0.5cm 1cm 1cm, clip]{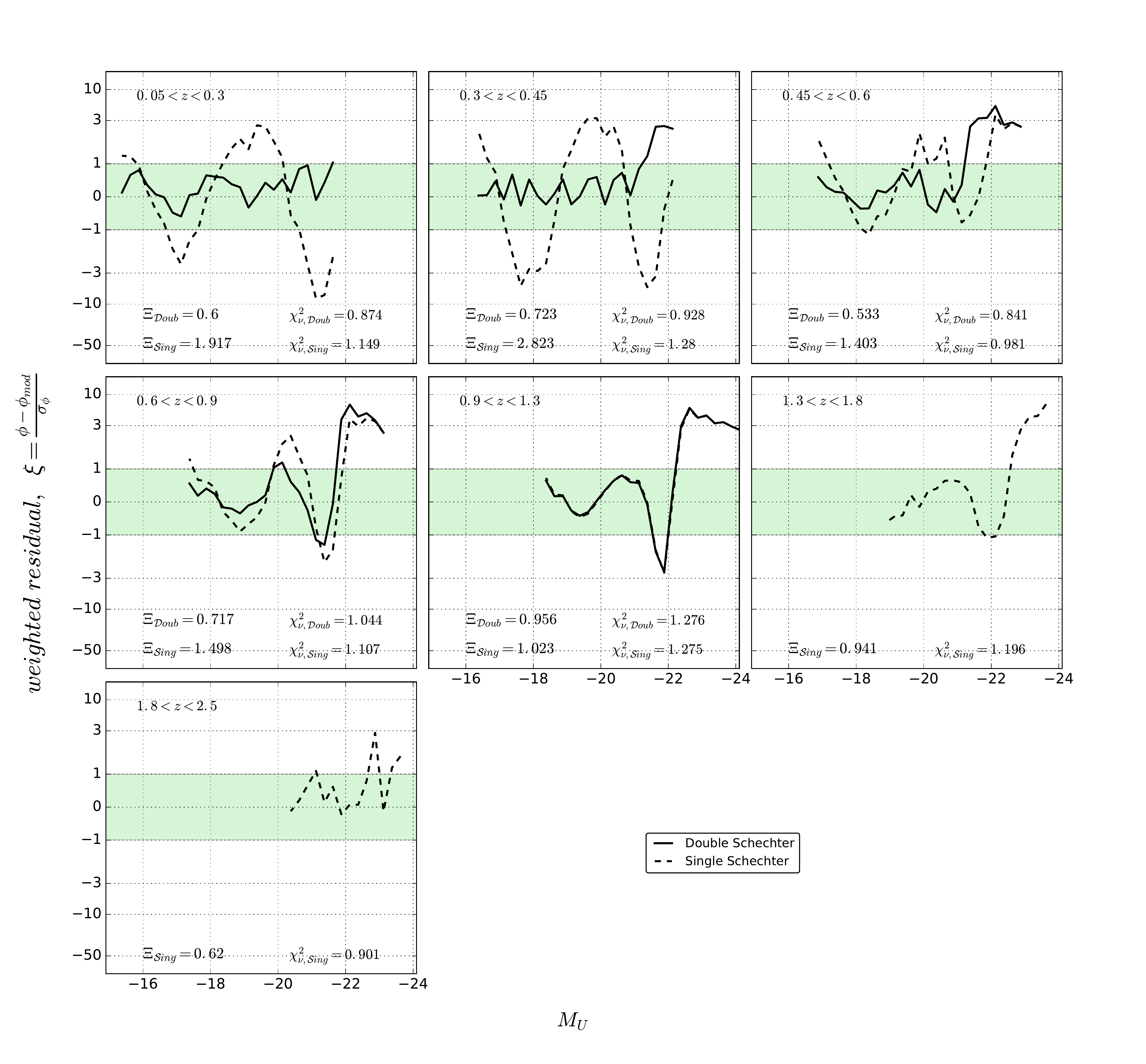}
\caption{
Weighted residuals $\xi$ as a function as a function of the U-band absolute magnitude: for our single-Schechter (dashed lines) and double-Schechter (solid lines) models of the U-band luminosity function. 
The green shaded area shows the region where the agreement between the model and the data is optimal (i.e., the residual is within the uncertainty), with $\vert \ \xi \ \vert = \vert \  \frac{\phi-\phi_{mod}}{\sigma_{\phi}} \ \vert \leq 1$. The associated typical deviation of the model (see Equation~\ref{eq_dev_mod}), $\Xi_\Sing$ and $\Xi_\Doub$, are reported in the lower-left corner of the sub-panels. 
While not optimal to compare the goodness-of-fit of non-linear models with different numbers of free parameters (see App. \ref{app_U_LF_fitting}), the corresponding reduced $\chi^2$ values are noted in the lower-right corners, for information.
\label{fig_U_LF_fit}  }
\end{figure*}

As observed in Figs. \ref{fig_U_LFs} and \ref{fig_field_U_LFs}, the U-band LF exhibits an upturn at the faint end, which results in a deviation from the shape of a pure Schechter function and argues for fitting the U-band LF with a double-Schechter function.

In order to assess whether a double-Schechter function is quantitatively better adapted to the U-band LF, we need to compare the the goodness-of-fit of two non-linear models with different numbers of parameters. While one might be inclined to compare the associated reduced $\chi^2$, defined as $\chi^2_\nu = \chi^2 / \nu$ (where $\nu$ is the number of degrees of freedom), $\nu$ is generally not of the commonly assumed form $\nu = N - M$ for non-linear models \citep[][]{Andrae2010}.  Consequently, the best way to compare the goodness-of-fit of single- and double-Schechter functions is actually to return to the distribution of the fit residuals.

To better appreciate the significance of the residual between the observed LF, $\phi$, and its parametric form, $\phi_{mod}$, it is relevant to consider the weighted residual, $\xi$, defined by
\begin{equation}
\label{eq_w_res}
\xi = \frac{\phi-\phi_{mod}}{\sigma_{\phi}} ~, 
\end{equation}
where the residual $\phi-\phi_{mod}$ is normalized by the LF uncertainty $\sigma_{\phi}$. 
Thereby, at given absolute magnitude, good agreement between the model and the data is met when the residual is smaller than the statistical uncertainty, i.e., when $-1 \lesssim \xi \lesssim 1$ (or when $\vert \ \xi \ \vert \lesssim 1$). Then, aiming at characterising the distribution of the residuals, it may be convenient to define $\Xi_{mod}$ as the normalized median absolute deviation of the model $\phi_{mod}$:
\begin{eqnarray}
\label{eq_dev_mod}
\Xi_{mod}  = 1.48 \times \mathrm{median}\left( \ \left\vert \frac{\phi-\phi_{mod}}{\sigma_{\phi}} \right\vert \ \right) \nonumber\\
= 1.48 \times  \mathrm{median}( \ \vert \ \xi \ \vert \ ) ~, ~~~~~~~~~~
\end{eqnarray}
$\Xi_{mod}$ being thereby a measure of the typical deviation of the model around the data relative to the statistical uncertainty on the data. In other words, there is overall good agreement of the model with the data when $\Xi_{mod} \lesssim 1$ and the better agreement, the smaller $\Xi_{mod}$ is.

In Fig. \ref{fig_U_LF_fit}, we plotted the weighted residual as a function of the U-band absolute magnitude, $M_U$, and we compare the residuals we obtained when fitting the LF with single- and double-Schechter functions. 
The typical deviation for the single- and double- Schechter fits, $\Xi_\Sing$ and $\Xi_\Doub$, are reported in the lower-left corner of each sub-panel in Fig. \ref{fig_U_LF_fit}, while the corresponding reduced $\chi^2$ are reported in the lower-right corners of the sub-panels, for information.

As one can see, the advantage of using a double-Schechter function to fit the U-band LF is clear up to $z = 0.9$. Associated residuals are indeed smaller than the statistical uncertainty from the faint end (notably around $M_U \sim -17$ where the upturn is observed; see Fig. \ref{fig_U_LFs}) to the bright end, before the disagreement start increasing around $M_U \sim -21.5$ due to Eddington bias (see Sect. \ref{sect_fit_Edd_treat}) and contamination by stars and quasars (see App. \ref{app_LF_sourcetype}).
One may notice that this translates into  $\Xi_\Doub < 0.75$, which traces a pretty good agreement between the data and the best-fit solution with a double-Schechter function, while the best single-Schechter solution is  clearly worse, with $\Xi_\Sing > 1.4$ (i.e., $\Xi_\Sing \simeq$ 2--4 $ \times \ \Xi_\Doub$).
At $0.9 < z < 1.3$, although single- and double-Schechter functions appear to provide similar results, the typical deviations of the two models tend to confirm that a double-Schechter profile better fits the U-band LF, with $\Xi_\Doub < 1 < \Xi_\Sing$. At higher redshifts, the completeness limits of our data ($M_{U,\ lim} = -18.85$ and $-20.32$ at $1.3 < z < 1.8$ and  $1.8 < z < 2.5$, respectively) prevent us from detecting any excess of galaxies at the faint end (the excess is typically visible for $M_U \gtrsim -17$ at lower redshift, as recalled above).  No definitive conclusion can therefore be drawn about the relevance of fitting the U-band LF with a double-Schechter function at $z > 1.3$, where a simple Schechter function fits well the LF (for $M_U < M_{U,\ lim}$), with $\Xi_\Sing < 1$ .

\section{Variation of the luminosity functions from field to field}
\label{app_LF_perfield}

\begin{figure*}
\center
\includegraphics[width=\hsize, trim = 6cm 6cm 7cm 6cm, clip]{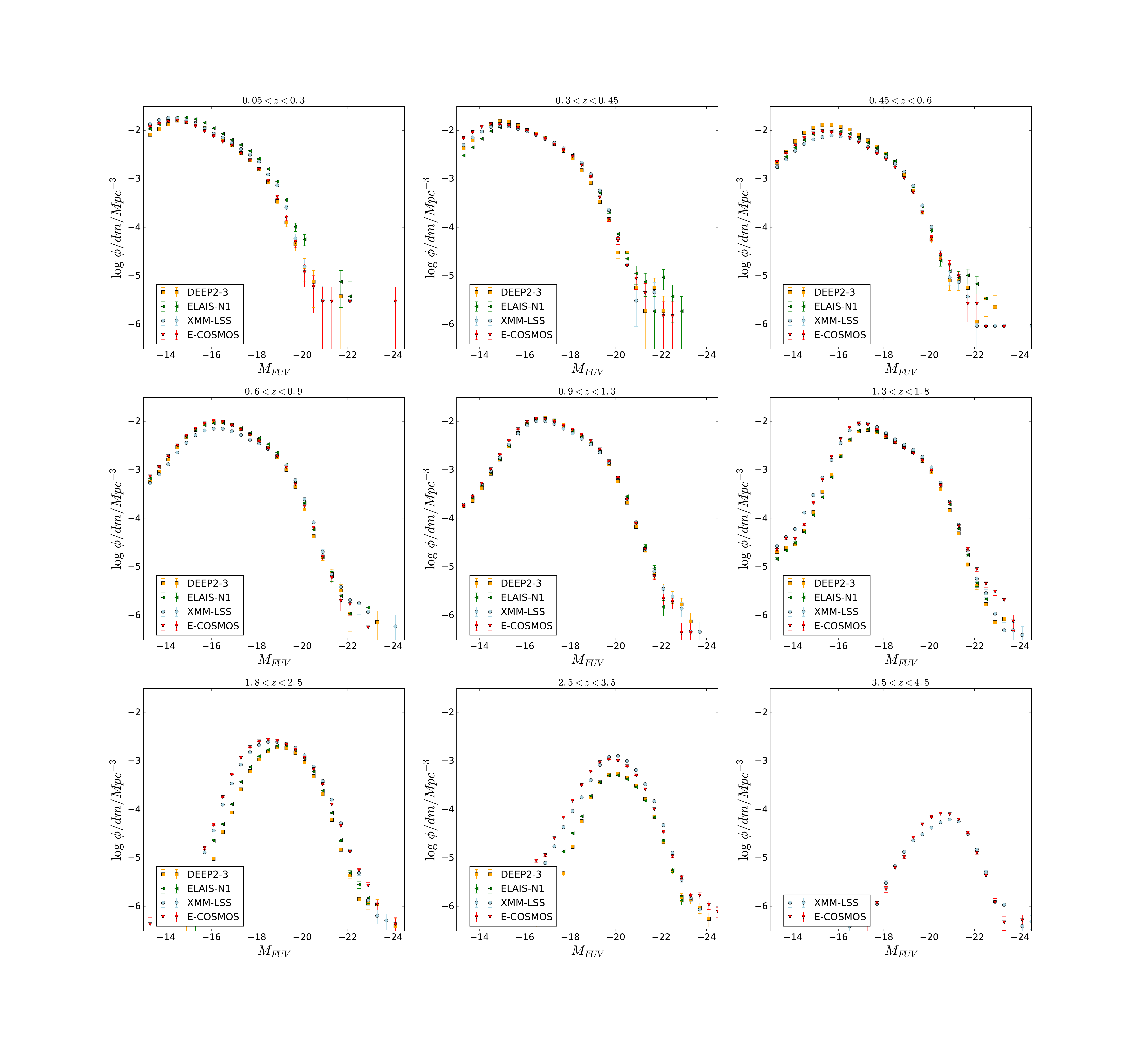}
\caption{FUV luminosity function in the four fields of CLAUDS+HSC: DEEP2-3, ELAIS-N1, XMM-LSS and E-COSMOS.
\label{fig_field_FUV_LFs}  }
\end{figure*}

\begin{figure*}
\center
\includegraphics[width=\hsize, trim = 6cm 6cm 7cm 6cm, clip]{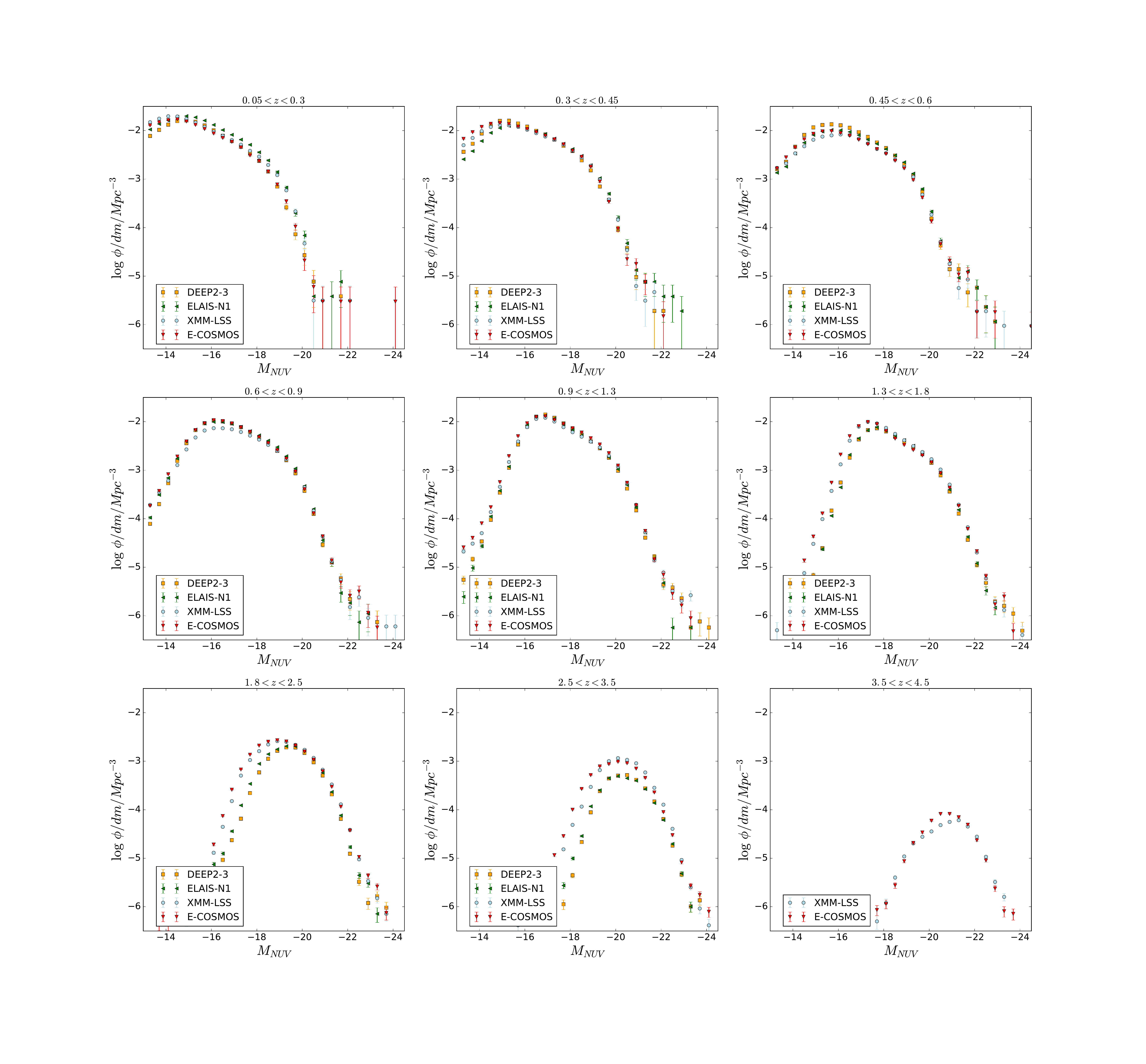}
\caption{NUV luminosity function in the four fields of CLAUDS+HSC: DEEP2-3, ELAIS-N1, XMM-LSS and E-COSMOS.
\label{fig_field_NUV_LFs}  }
\end{figure*}

\begin{figure*}
\center
\includegraphics[width=\hsize, trim = 6cm 6cm 7cm 6cm, clip]{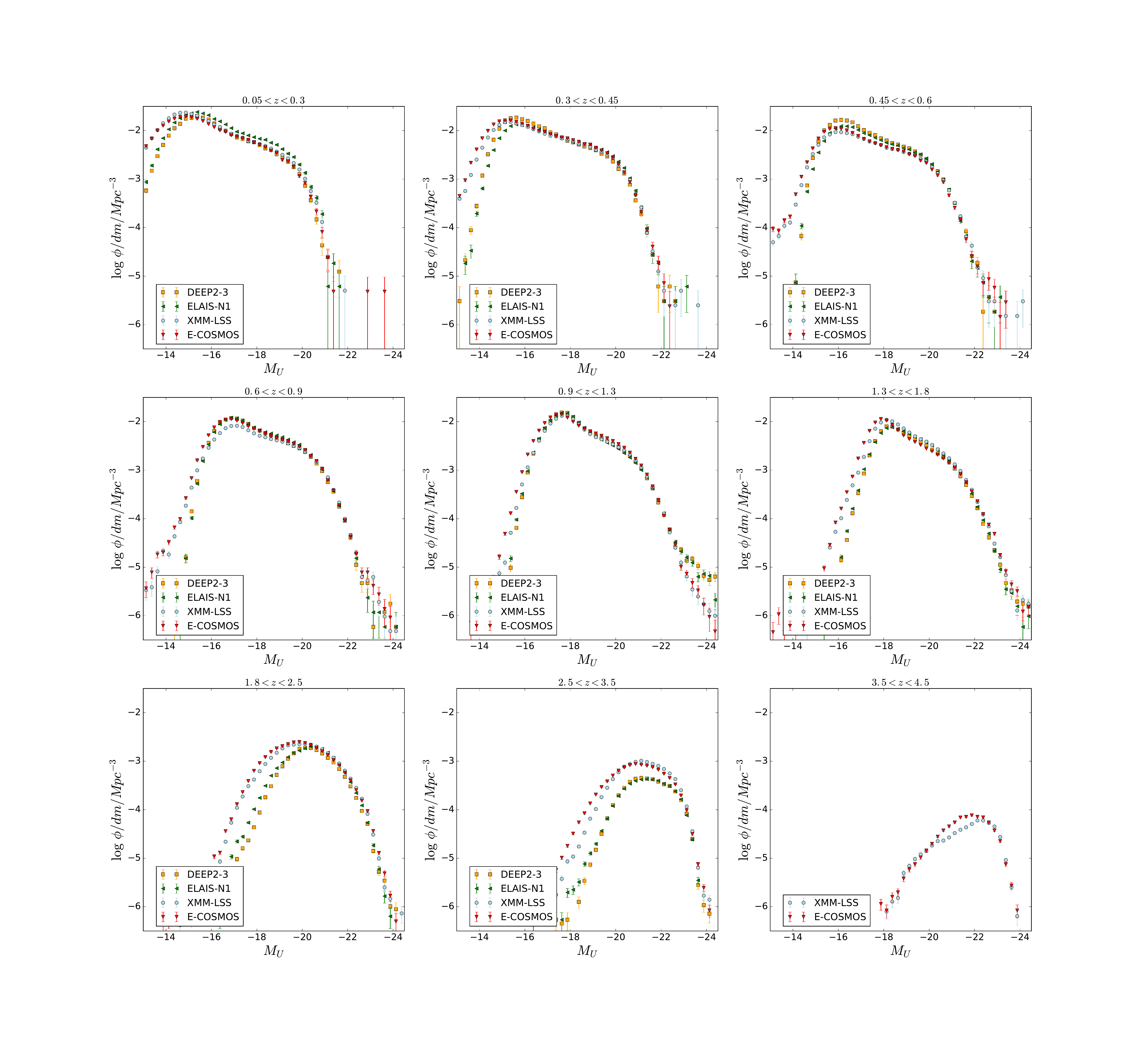}
\caption{U-band luminosity function in the four fields of CLAUDS+HSC: DEEP2-3, ELAIS-N1, XMM-LSS and E-COSMOS.
\label{fig_field_U_LFs}  }
\end{figure*}

Figures \ref{fig_field_FUV_LFs}, \ref{fig_field_NUV_LFs} and \ref{fig_field_U_LFs} show, respectively, the FUV, NUV and U-band raw LFs we measured in our eight redshift bins, for each of the four fields of our survey: DEEP2-3, ELAIS-N1, XMM-LSS and E-COSMOS.
In these figures, the Deep and Ultra-Deep layers are not separated, which explains why XMM-LSS and E-COSMOS, which contain the Ultra-Deep layer, appear deeper than DEEP2-3 and ELAIS-N1.

The deviation between the LFs measured in each field illustrates the cosmic variance affecting the LF measurement in each field. One can see how the cosmic variance depends on the cosmic volume probed for a given redshift bin and a given effective area: it is thereby not surprising to observe the largest deviation between field LFs in our lowest redshift bin, $0.05 < z \leq 0.3$.

On the other hand, the deviation between the LFs one can observe from field to field at the extremely bright end, for very small comoving densities $<10^{-5}$ Mpc$^{-3}$, is likely to be due contamination by stars and QSOs that could not be discarded by the procedure described in Sect. \ref{sect_photoz}. Indeed, the photometric identification of stars and QSOs depends on the SNR of the sources (i.e., the depth of the data), which is different in our different fields. In that respect, what can be seen in Fig. \ref{fig_field_U_LFs} at $0.9 < z \leq 1.3$ (where most of the U-band absolute magnitudes are derived from observed $i$-band) is particularly striking but not surprising: XMM-LSS and E-COSMOS host indeed our Ultra-Deep layer and include much deeper $g,r,i,z,y$ observations than DEEP2-3 and ELAIS-N1.

\section{Luminosity function per type of source}
\label{app_LF_sourcetype}

\begin{figure*}
\center
\includegraphics[width=\hsize, trim = 6cm 6cm 7cm 6cm, clip]{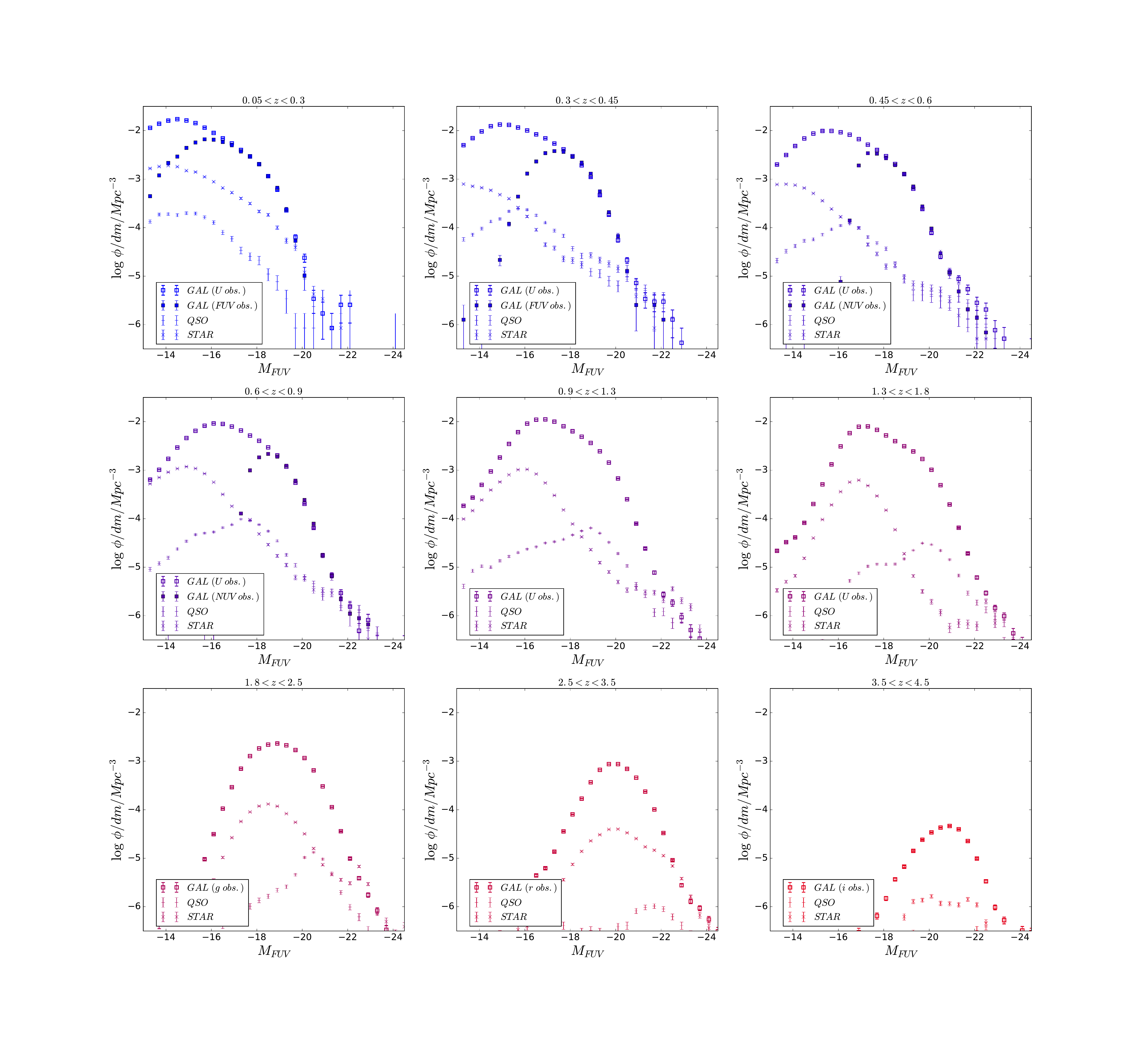}
\caption{FUV luminosity function of the different types of sources identified in our analysis: galaxies (GAL), quasars (QSO) and stars. The very bright end is mostly populated by stars and QSOs, which confirms that the deviation of the bright end of galaxy FUV luminosity function from a pure Schechter form is likely to be due to contamination from stars and QSOs.
\label{fig_type_FUV_LFs}  }
\end{figure*}

\begin{figure*}
\center
\includegraphics[width=\hsize, trim = 6cm 6cm 7cm 6cm, clip]{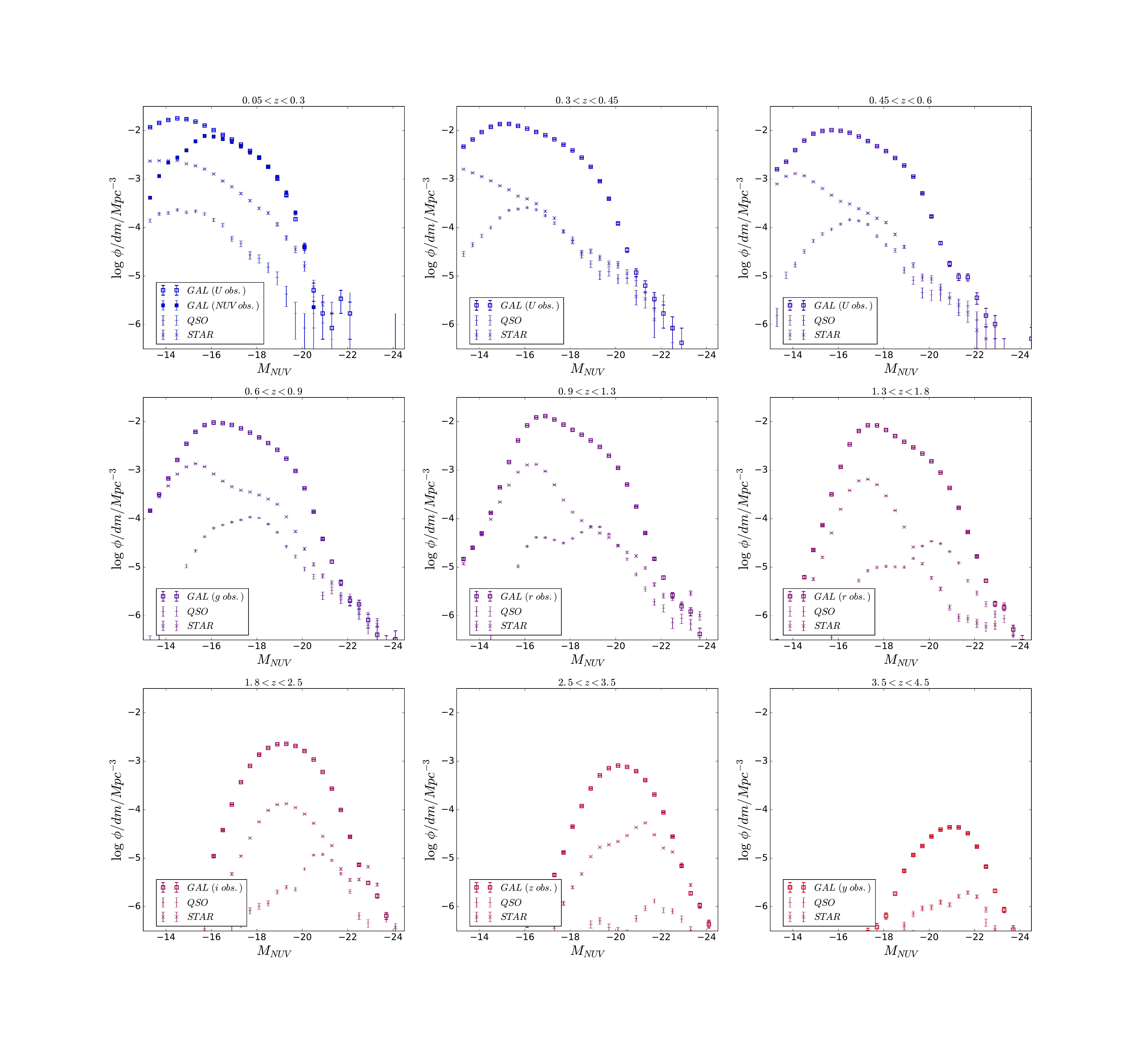}
\caption{NUV luminosity function of the different types of sources identified in our analysis: galaxies (GAL), quasars (QSO) and stars. The very bright end is mostly populated by stars and QSOs, which confirms that the deviation of the bright end of galaxy NUV luminosity function from a pure Schechter form is likely to be due to contamination from stars and QSOs.
\label{fig_type_NUV_LFs}  }
\end{figure*}

\begin{figure*}
\center
\includegraphics[width=\hsize, trim = 6cm 6cm 7cm 6cm, clip]{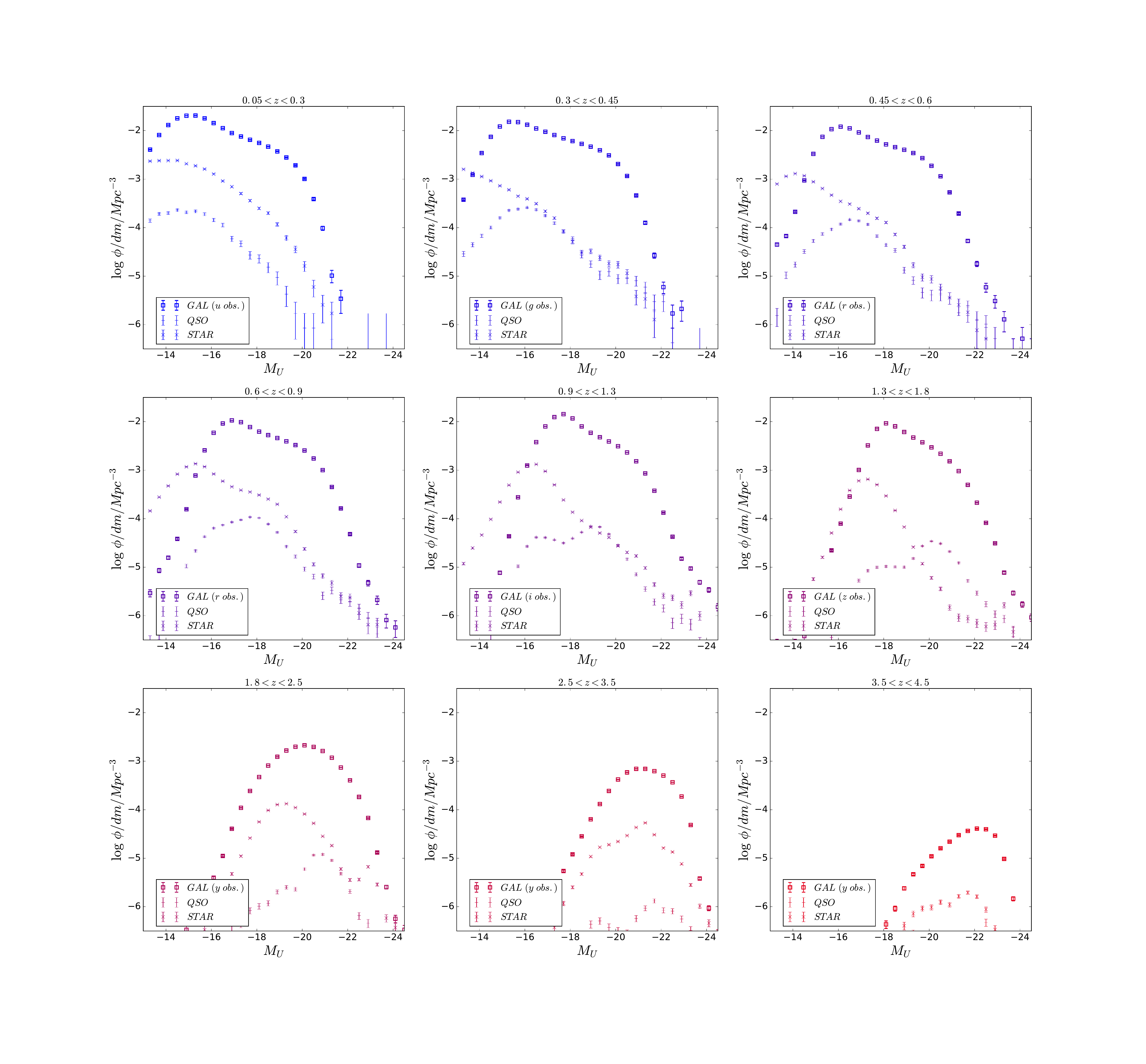}
\caption{U-band luminosity function of the different types of sources identified in our analysis: galaxies (GAL), quasars (QSO) and stars. The very bright end is mostly populated by stars and QSOs, which confirms that the deviation of the bright end of galaxy U-band luminosity function from a pure Schechter form is likely to be due to contamination from stars and QSOs.
\label{fig_type_U_LFs}  }
\end{figure*}

Figures \ref{fig_type_FUV_LFs}, \ref{fig_type_NUV_LFs} and \ref{fig_type_U_LFs} show, respectively, the FUV, NUV and U-band LFs we could measure in our eight redshift bins, depending on the type of sources we identified with the procedure described in Sect. \ref{sect_photoz}: galaxies, quasars and stars. 

Stars with $z > 0$ are obviously not real, and the redshifts of QSOs are most probably wrong, but the exercise allows us to see how and where these two populations may contaminate our LF measurements. Indeed, as one can see in all FUV, NUV and U-band LFs, sources classified as stars and QSOs are completely dominated by the galaxy population at low luminosities (faint absolute magnitudes) down to the completeness limit, but their incidence increases with increasing luminosity to become as numerous as galaxies at the very bright end of the LFs.

This seems to confirm that the extremely bright end of our LF measurements may significantly suffer from contamination by stars and QSOs, typically for comoving densities $< 10^{-5}$ Mpc$^{-3}$. None the less, we verified that this very limited population did not affect our analysis, as described in Sect. \ref{sect_fit_Edd_treat}.

%


\bsp	
\label{lastpage}

\end{document}